\documentclass[fleqn,usenatbib]{mnras}

\usepackage{newtxtext,newtxmath}

\usepackage[T1]{fontenc}
\usepackage{ae,aecompl}

\usepackage{graphicx}	
\usepackage{amsmath}	
\usepackage{amssymb}	
\usepackage{subfig}	

\newcommand{\ha}{H$\alpha$~}
\newcommand{\hb}{H$\beta$~}
\newcommand{\hans}{H$\alpha$}
\newcommand{\hbns}{H$\beta$}
\def \ll {$\lambda\lambda$}
\def \l {$\lambda$}
\def \kms {km\,s$^{-1}$}
\newcommand{\ergs}{\,erg\,s$^{-1}$} 

\title{Kinematics of \ion{C}{\sc  IV} and [\ion{O}{\sc III}] emission in luminous high-redshift quasars}

\author [L. Coatman et al.]{
Liam Coatman,$^{1}$
Paul C. Hewett,$^{1}$\thanks{E-mail: phewett@ast.cam.ac.uk}
Manda Banerji,$^{1,2}$ 
Gordon T. Richards,$^{3}$ \newauthor
Joseph F. Hennawi$^{4,5}$ and
J. Xavier Prochaska$^{6}$
\\
$^{1}$Institute of Astronomy, University of Cambridge, Madingley Road, Cambridge, CB3 0HA, UK\\
$^{2}$Kavli Institute for Cosmology, University of Cambridge, Madingley Road, Cambridge, CB3 0HA, UK\\
$^{3}$Department of Physics, Drexel University, 3141 Chestnut Street, Philadelphia, PA 19104, USA\\
$^{4}$Department of Physics, Broida Hall, University of California, Santa Barbara, CA 93106-9530, USA\\
$^{5}$Max-Planck-Institut f{\"u}r Astronomie, K{\"o}nigstuhl 17, D-69117 Heidelberg, Germany\\
$^{6}$UCO/Lick, University of California, Santa Cruz, CA 95064, USA
}
\date{Accepted XXX. Received YYY; in original form ZZZ}

\pubyear{2018}

\begin{document}
\label{firstpage}
\pagerange{\pageref{firstpage}--\pageref{lastpage}}
\maketitle

\begin{abstract}
We characterise ionised gas outflows using a large sample of $\simeq$330 high-luminosity ($45.5 <$ log(L$_{\rm{bol}}$/erg s$^{-1}$) $<49.0$), high-redshift (1.5$ \lesssim$z$\lesssim$4.0) quasars via their [\ion{O}{III}]\ll4960,5008 emission. The median velocity width of the [\ion{O}{III}] emission line is 1540 kms$^{-1}$, increasing with increasing quasar luminosity. Broad, blue-shifted wings are seen in the [\ion{O}{III}] profiles of $\simeq$42 per cent of the sample. Rest-frame ultraviolet spectra with well-characterised \ion{C}{IV}\l1550 emission line properties are available for more than 210 quasars, allowing an investigation  of the relationship between the Broad Line Region (BLR) and Narrow Line Region (NLR) emission properties. The [\ion{O}{III}] blueshift is correlated with \ion{C}{IV} blueshift, even when the dependence of both quantities on quasar luminosity has been taken into account. A strong anti-correlation between the [\ion{O}{III}] equivalent width (EW) and \ion{C}{IV} blueshift also exists. Furthermore, [\ion{O}{III}] is very weak, with EW$<$1\,\AA\, in $\simeq$10 per cent of the sample, a factor of 10 higher compared to quasars at lower luminosities and redshifts. If the [\ion{O}{III}] emission originates in an extended NLR, the observations suggest that quasar-driven winds are capable of influencing the host-galaxy environment out to kilo-parsec scales. The mean kinetic power of the ionised gas outflows is then 10$^{44.7}$ erg s$^{-1}$, which is $\simeq$0.15 per cent of the bolometric luminosity of the quasar. These outflow efficiencies are broadly consistent with those invoked in current active galactic nuclei feedback models.

\end{abstract}




\section{Introduction}

X-ray and ultra-violet spectroscopy has revealed high-velocity outflows to be nearly ubiquitous in high-accretion-rate quasars \citep[e.g.][]{tombesi10,ganguly08}.
Strong evidence for high-velocity outflows in the vicinity of quasars include Broad Absorption Lines (BALs), Narrow Absorption Lines (NALs) and blueshifted emission-lines. 
These observations suggest that the energy released by quasars can have a dramatic effect on their immediate environments. 

Quasars driving powerful outflows over galactic scales has become a central tenet of galaxy evolution models involving `quasar feedback' \citep[e.g.][]{silk98,springel05,bower06}.
In recent years, significant resource has been devoted to searching for observational evidence of the phenomenon, leading to
detections of outflows in quasar-host galaxies using tracers of atomic, molecular, and ionised gas \citep[e.g.][]{nesvadba06,arav08,nesvadba08,moe09,alexander10,dunn10,feruglio10,nesvadba10,alatalo11,cano-diaz12,harrison12,harrison14,cimatti13,rupke13,veilleux13,cicone14,nardini15}.  

One particularly successful technique has been to use forbidden quasar emission-lines to probe the dynamical state of the ionised gas extended over kilo-parsec scales in quasar host-galaxies. 
Because of its high equivalent width, [\ion{O}{III}]\l\l4960,5008\footnote{Vacuum wavelengths are employed throughout the paper.} is the most studied of the narrow line region (NLR) emission-lines. 
The [\ion{O}{III}] emission is found to consist of two distinct components: a `core' component, with a velocity close to the systemic redshift of the host-galaxy, and a broader `wing' component, which is asymmetric and normally extends blueward. 
The core component velocity-width includes a contribution from the gravitational potential of the host-galaxy and the central supermassive black hole.
However, the velocity-width of the blue wing is too broad for the gas to be in dynamical equilibrium with the host-galaxy \citep[e.g.][]{liu13}. The general consensus is that the blueshifted [\ion{O}{III}]-emission traces relatively high-velocity outflowing gas. The receding side of the outflow may be obscured due to the presence of dust, either in the outflow or elsewhere, and only the near-side of the outflow, which is blueshifted, is observed. 
The relative balance between the core and wing components varies significantly from object to object, and governs the width and asymmetry of the overall [\ion{O}{III}] emission profile \citep[e.g.][]{shen14}. 

Observations of broad velocity-widths and blueshifts in narrow emission-lines stretch back several decades \citep[e.g.][]{weedman70,stockton76,heckman81,veron81,feldman82,heckman84,vrtilek85,whittle85,boroson92}. 
These studies rely, however, on small samples, which may target particular sub-types of active galactic nuclei (AGN), which are not representative of the properties of the quasar population as a whole. 
More recently, the advent of large optical spectroscopic surveys \citep[e.g. the Sloan Digital Sky Survey; SDSS;][]{york00} has facilitated studies of the NLR in tens of thousands of quasars \citep[e.g.][]{boroson05,greene05a,zhang11,mullaney13,zakamska14,shen14}, and the population statistics have provided constraints on the prevalence and drivers of ionised outflows.   
At the same time, spatially resolved spectroscopy has revealed that at least a fraction of these outflows are extended over galaxy scales \citep[e.g.][]{greene09,greene11,harrison12,hainline13,harrison14,bae17} although \citet{karouzos16} caution regarding some of the inferred large spatial extents of the outflows.

In general, however, these studies are based on AGN at relatively low redshifts, $z\lesssim0.8$, where the [\ion{O}{III}] emission is present in optical spectra. Investigations of large samples have not included the redshift range when star formation and supermassive black hole (BH) accretion peaked ($2 \lesssim z \lesssim 4$), which is when correlations between properties of the galaxy bulge and BH (e.g. the stellar velocity dispersion - BH mass relation) are thought to have been established. 
At these redshifts bright optical emission-lines including the [\ion{O}{III}] doublet are redshifted to near-infrared wavelengths, where observations are far more challenging. 
As a consequence, observations at high redshifts have typically relied on relatively small numbers of objects.
These studies find [\ion{O}{III}] to be broader in more luminous quasars, with velocity-widths $\gtrsim1000$\,\kms common \citep[e.g.][]{netzer04,kim13,brusa15,shen16a}.  
These findings suggest that quasar efficiency in driving galaxy-wide outflows increases with luminosity \citep[e.g.][]{netzer04,nesvadba08,kim13,brusa15,carniani15,perna15,bischetti16}. 
The fraction of objects with very weak [\ion{O}{III}] emission also appears to increase with redshift and/or luminosity \citep[e.g.][]{netzer04}. 

In this paper, we analyse the [\ion{O}{III}] properties of a sample of 354 high-luminosity, redshift $1.5 \lesssim z \lesssim 4.0$ quasars selected from the near-infrared spectroscopic catalogue presented in \citet{coatman17}.
To date, this is the largest study of the NLR-properties of high-redshift, high-luminosity, broad-line quasars. 
In Section \ref{sec:data} we describe our quasar sample and spectral line measurements.
Section \ref{sec:results} presents our main results on the [\ion{O}{III}] line strength and kinematics observed in the sample while Section \ref{sec:discussion} discusses the potential implications of the observational results. 

\section{Data}

\label{sec:data}

\subsection{Quasar sample}

From our near-infrared spectroscopic catalogue \citep{coatman16, coatman17}, we have selected 354 quasars which have spectra covering the [\ion{O}{III}]\l\l4960.30,5008.24 doublet. 
The broad Balmer \hb emission-line (4862.68\,\AA) has also been observed for all but two of the sample. 
For 165 quasars, the spectra extend to cover the broad \ha emission-line (6564.61\,\AA).
Optical spectra, the majority from the SDSS, including the rest-frame ultraviolet \ion{C}{IV}\l\l1548,1550 emission are available for 258 objects.
The distribution of the quasars in the redshift-luminosity plane is discussed in \citet{coatman17} but, in summary, the sample covers a wide range in redshift ($1.5 \lesssim z \lesssim 4$) and luminosity ($45.5 \lesssim \log L_{\text{bol}} \lesssim 49.0$). 
The spectrographs and telescopes used to obtain the near-infrared spectra are summarised in Table~\ref{tab:specnums_ch4}.

\begin{table}
  \centering
  \footnotesize 
    \begin{tabular}{ccc} 
    \hline
    Spectrograph & Telescope & Number \\
    \hline
    FIRE         & MAGELLAN  & 31 \\
    GNIRS        & GEMINI-N  & 28 \\
    ISAAC        & VLT       & 7 \\
    LIRIS        & WHT       & 7 \\
    NIRI         & GEMINI-N  & 29 \\
    NIRSPEC      & Keck II   & 3 \\
    SINFONI      & VLT       & 80 \\
    SOFI         & NTT       & 76 \\
    TRIPLESPEC   & ARC-3.5m  & 27 \\
    TRIPLESPEC   & P200      & 45 \\
    XSHOOTER     & VLT       & 21 \\
    \hline
    \multicolumn{2}{c}{Total} & 354 \\
    \hline
    \end{tabular}
    \caption[{The numbers of quasars with [\ion{O}{III}\l\l4960,5008] line measurements and the spectrographs and telescopes used to obtain the near-infrared spectra.}]{The numbers of quasars with [\ion{O}{III}] line measurements and the spectrographs and telescopes used to obtain the near-infrared spectra.}
  \label{tab:specnums_ch4}
\end{table} 

\subsection{Spectral Line Measurements}
\label{sec:line_measurements}

The data reduction and procedures employed to derive emission-line properties are described in \citet{coatman17}. 
\ion{C}{IV} emission-line properties (used to infer the strength of BLR outflows) are taken directly from \citet{coatman16, coatman17}.  
Here we review the main elements of the scheme used to obtain parameters describing the H$\alpha$, H$\beta$ and [\ion{O}{III}] emission properties.
Our approach is to model the \hbns/[\ion{O}{III}] complex using a power-law continuum, an empirical \ion{Fe}{II} template (taken from \citealt{boroson92}) and multiple Gaussian components.
Non-parametric properties are then derived from the best-fitting model. 
Compared to measuring properties directly from the data, the approach is more robust when analysing spectra with limited signal-to-noise ratio (S/N) and more effective when estimating properties of partially blended emission-lines.
Shen and collaborators have established a careful and well-defined approach to such a parametrization of rest-frame optical emission lines using near-infrared spectra of quasars \citep[e.g.][]{shen11,shen12,shen16a} and we have deliberately ensured that our analysis follows their approach closely. For consistency we also adopt the 
\citet{shen12} definition of a full-width at half-maximum (FWHM) =1200\,\kms as the division between emission ascribed to the `narrow' and `broad' emission-line regions. 
Adopting FWHM=1200\kms as the boundary is a pragmatic choice that breaks the potential degeneracy between `broad' and `narrow' emission-line regions for the majority of objects. The absolute value of the FWHM adopted is, to a degree, arbitrary and no real significance is ascribed to the properties of the small number of objects with emission-line profiles where the FWHM lies close to the boundary.

Before a spectrum can be modelled, it must first be transformed to the rest-frame of the quasar.  
The redshift used in this transformation is either derived from the peak of the broad \ha emission ($\simeq$40 per cent of our sample), from the peak of the broad \hb emission ($\simeq$40 per cent) or from the peak of the narrow [\ion{O}{III}] emission ($\simeq$20 per cent).
The rest-frame transformation is only required to be accurate to within $\simeq$1000\,\kms\, of the true systemic redshift for our fitting procedure to work. 
In Appendix \ref{sec:ch4_redshifts}, we compare more precise systemic redshift estimates based on our parametric model fits to the [\ion{O}{III}], \hb and \ha lines, and demonstrate that there is no significant systematic offset between the results using the three different emission lines. 
The scatter between the different redshift estimates is $\sim$300\,\kms, much smaller than the outflow velocities we derive later in the paper. 
The results of our study are therefore essentially unaffected by the choice of systemic redshift.
Thus, the estimation of systemic redshifts using the hydrogen Balmer lines and the [\ion{O}{III}] lines is essentially straightforward. 
It is worth noting, however, that such is not the case when only rest-frame ultraviolet spectra are available and considerable care should be taken to understand the substantial systematic errors in the calculation of systemic redshifts as a function of the intrinsic quasar ultraviolet SED \cite[e.g.][]{hewett10}.

\begin{figure*}
    \centering
    \includegraphics[width=\textwidth]{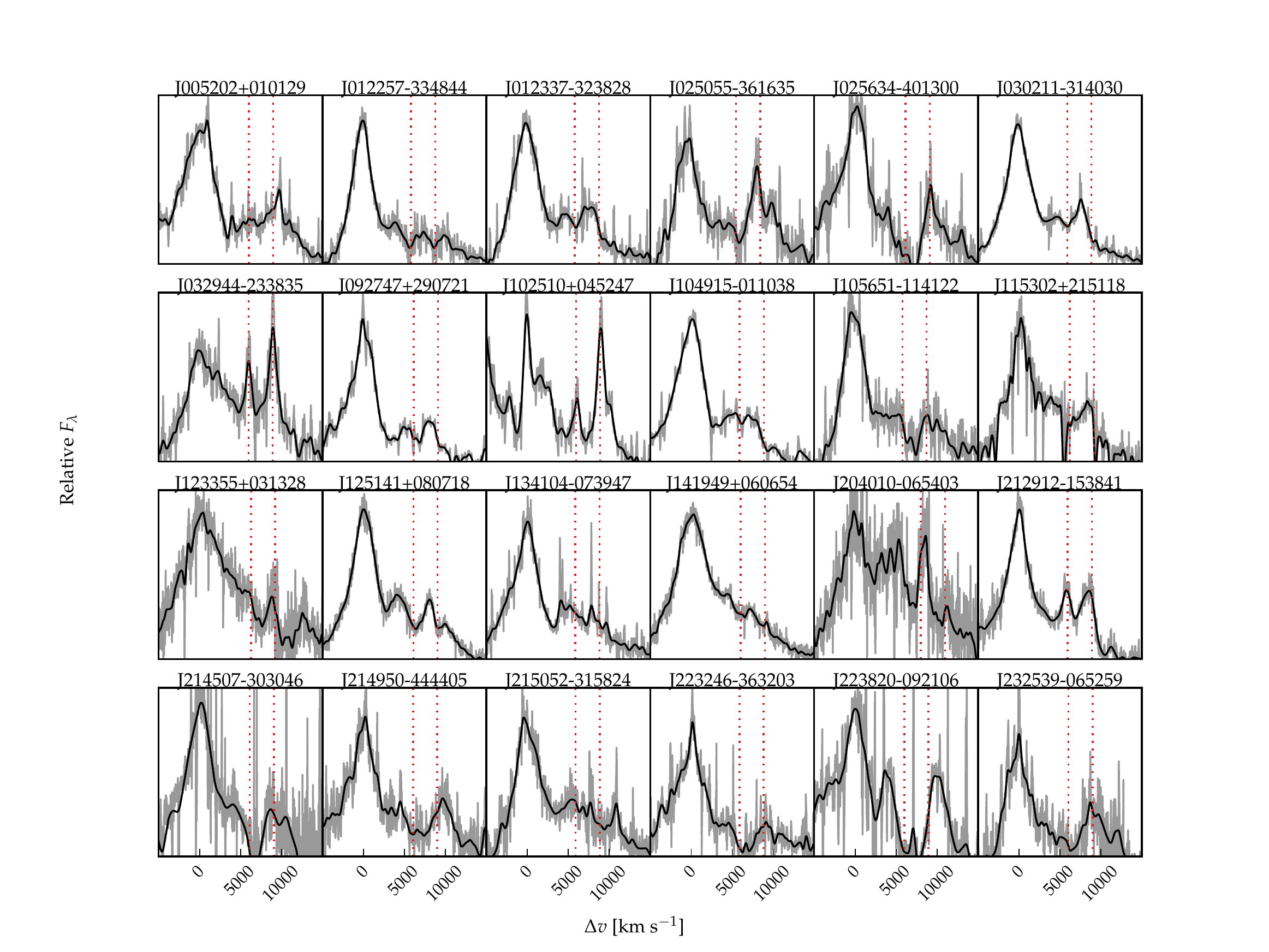} 
    \caption{Spectra of the 24 objects for which significant \ion{Fe}{II} emission is still present following our \ion{Fe}{II}-subtraction procedure. The original spectra are shown in grey with versions smoothed via convolution with a 200\,\kms\, Gaussian kernel shown in black. The vertical lines indicate the expected positions of the [\ion{O}{III}]\l\l4960,5008 doublet (which is generally very weak in these objects) with the systemic redshift defined using the peak of the broad \hb emission. Significant residual \ion{Fe}{II} emission is present over some 10\,000-15\,000\kms of the 20\,000\kms-wide interval plotted for each spectrum.}     
    \label{fig:bad_fe}
\end{figure*}

Prior to modelling the \hbns/[\ion{O}{III}] emission, we first subtract the nearby continuum and \ion{Fe}{II} emission. 
A power-law continuum and an optical \ion{Fe}{II} template - taken from \citet{boroson92} - are employed, using two `windows' at 4435-4700 and 5100-5535\,\AA. The \ion{Fe}{II}  template is convolved with a Gaussian, and the width of this Gaussian, along with the normalisation and velocity offset of the \ion{Fe}{II} template, are free variables in the pseudo-continuum fit. 
For 24 objects, the procedure failed to remove the \ion{Fe}{II} emission adequately from the spectra (Fig.~\ref{fig:bad_fe})
because the relative strengths of the \ion{Fe}{II} lines differ significantly from those of I Zw 1, on which the \citet{boroson92} \ion{Fe}{II} template is based. 
The residual \ion{Fe}{II} emission is at rest-frame wavelengths very close to the laboratory wavelengths of the [\ion{O}{III}] doublet, which is generally weak. 
As a result, the [\ion{O}{III}] parameters we derive are unreliable and the
24 objects are therefore excluded from our analysis (leaving 330 quasars). 

To illustrate the importance of the \ion{Fe}{II} subtraction procedure in order to measure [\ion{O}{III}]-emission properties reliably, we consider the object J223820-092106 (shown bottom row, second from right in Fig.~\ref{fig:bad_fe}). 
J223820-092106 was also analysed by \citet{shen16a} and represents the only quasar for which there is any significant disagreement regarding the [\ion{O}{III}] emission from the two analyses. \citet{shen16a} find an [\ion{O}{III}] detection but with a significance of only 2.0\,$\sigma$.
In our analysis, however, the emission is identified as poorly-subtracted \ion{Fe}{II}. Irrespective of which analysis is correct, the comparison illustrates the degree of caution that should be exercised when interpreting the statistics of [\ion{O}{III}] emission detections of low significance\footnote{It is worth noting that the ultra-violet-derived redshift for J223820-092106 from \citet{hewett10} is significantly in error and the value of $z=2.33$, measured using H$\beta$ emission from \citet{shen16a} and in this paper is correct.}

\subsubsection{Modelling \hb  and [O\,{\sc iii}]}
\label{sec:oiiimodel}

\begin{table}
  \centering
  \footnotesize 
    \begin{tabular}{cccc} 
    \hline
    Model & Components & Fix centroids? & Number \\
    \hline
    a & 1 BG  & N/A &  9 \\
    b & 2 BG & Yes &  39 \\
    c & 2 BG & No  &  295 \\
    d & 2 BG + 1 NG & No & 9 \\
    \hline
    \end{tabular}
    \caption{Summary of models used to fit the \hb emission, and the number of quasars to which each model is applied. The number of broad (BG) and narrow (NG) Gaussian components is given along with, when applicable, whether the centroids of the broad Gaussian components are constrained to be identical.}
  \label{tab:hbmod}
\end{table} 

In the majority of objects, broad \hb emission is modelled using two Gaussians with non-negative amplitudes and FWHM$>$1200\,\kms.
In nine objects \hb is modelled with a single Gaussian and in 39 objects \hb is modelled with two Gaussians, but the velocity centroids of the two Gaussians are constrained to be equal. 
The 48 spectra generally have low S/N, and adding extra freedom to the model does not significantly decrease the reduced-$\chi^2$.
In addition, there are cases when the blue wing of the \hb emission extends beyond the lower wavelength limit of the spectra; in these cases models with more freedom are insufficiently constrained by the data.    

\begin{figure*}
    \centering
    \includegraphics[width=\textwidth]{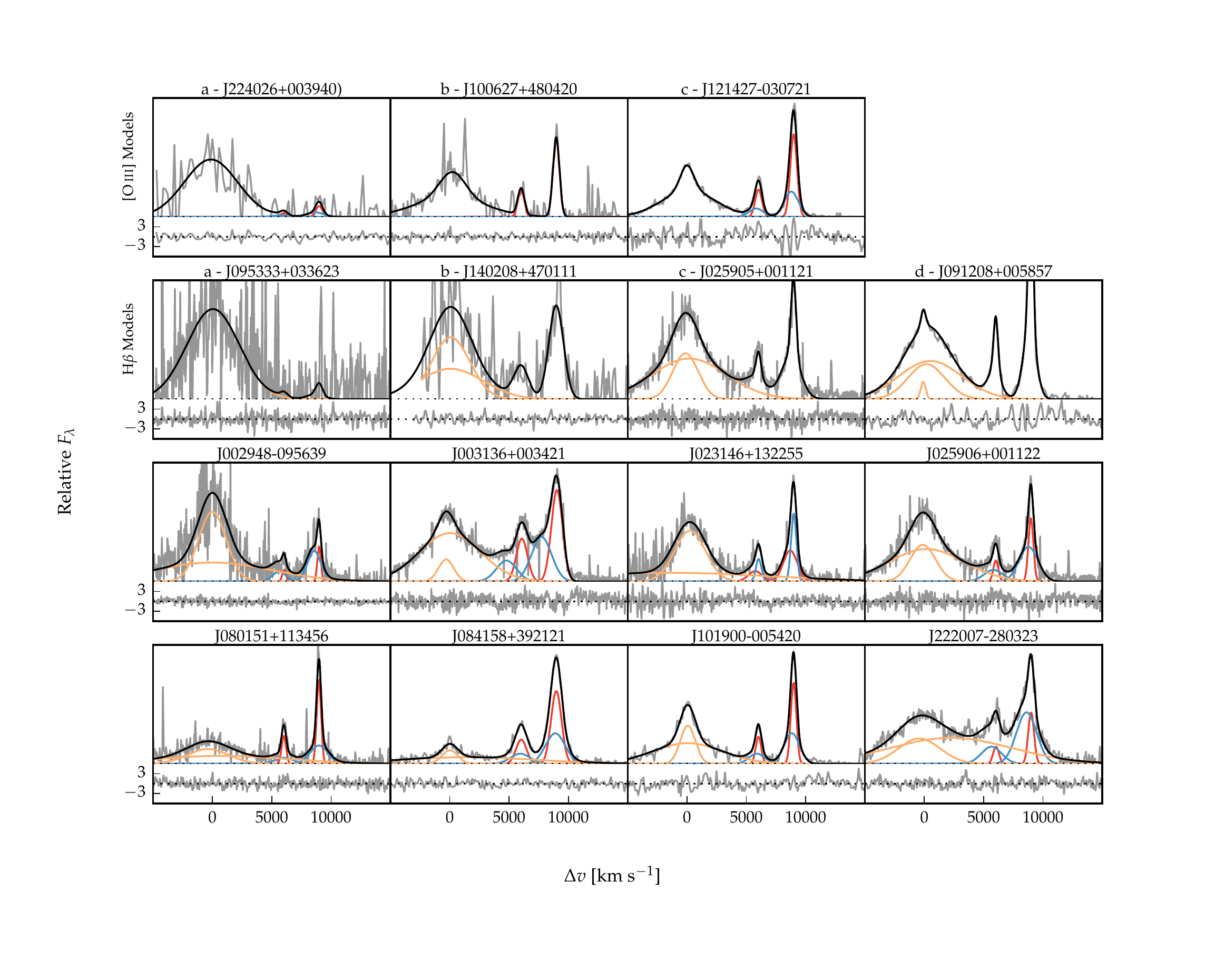} 
    \caption{Example model fits to the continuum- and \ion{Fe}{II}-subtracted \hbns/[\ion{O}{III}] emission lines. 
    The data is shown in grey, the best-fitting model in black, and the individual model components in orange. The peak of the [\ion{O}{III}]\l\l4960,5008 emission is used to set the redshift, and $\Delta{v}$ is the velocity shift from the rest-frame transition wavelength of \hbns. The data-minus-model residuals, scaled by the errors on the fluxes, are plotted below each spectrum. The top row shows examples of fits employing the three different [\ion{O}{III}] models (see Table~\ref{tab:oiiimod}).The second row shows examples of fits employing the four different H$\beta$ models (see Table~\ref{tab:hbmod}). The two bottom rows show fits to eight objects selected to illustrate the range of emission properties in the sample.}     
    \label{fig:example_spectrum_grid}
\end{figure*}

Contributions to the \hb emission from the NLR (FWHM $\lesssim$1200\kms) are generally weak, and an additional Gaussian component to model NLR-emission is not required. 
Features in the spectrum-minus-model residuals of nine spectra, however, indicate that a narrow emission component is significant, and an additional narrow Gaussian is therefore included. 
If a NLR-contribution to the \hb emission is present in more of the sample, then measures of the \hb velocity-width will be biased to lower values. 
Our systemic redshift estimates, however, that use the peak of the \hb emission (Appendix~\ref{sec:ch4_redshifts}), will not be affected. 
The type of \hb model, and the numbers of quasars to which each model is applied, are summarised in Table~\ref{tab:hbmod}. Example fits to spectra using the four models for H$\beta$ are shown in the second row of Fig.~\ref{fig:example_spectrum_grid}.

Each component of the [\ion{O}{III}]-doublet is fit with one or two Gaussians, depending on the fractional reduced-$\chi^2$ difference between the one- and two-Gaussian models.
The peak flux ratio of the Gaussians for the [\ion{O}{III}] 4960\,\AA\, and 5008\,\AA\, components are fixed at the expected 1:3 ratio and the width and velocity offsets are constrained to be equal\footnote{For J$003136$+$003421$, a significantly better fit ($\Delta \chi^2_{\nu} \sim 25$ per cent) is obtained when the peak flux ratio constraint is relaxed; the peak ratio of the best-fitting model is 1:2.13.}.
The first Gaussian, with FWHM constrained to be $\le$1200\,\kms, primarily models what might be termed the conventional NLR-emission contribution.
A second Gaussian, which may have FWHM$>$1200\,\kms, is included if the reduced-$\chi^2$ of the model decreases by more than 5 per cent.
In practice, the second Gaussian models the, normally blueward, asymmetric contribution to the [\ion{O}{III}]-doublet.
One hundred and twenty-eight spectra are fit with a single Gaussian and $140$ with two Gaussians. 

For the remaining 62 objects, with very weak [\ion{O}{III}] (mean equivalent width (\text{EW})$\simeq$2\,\AA), the Gaussian model is poorly constrained, leading to large errors on the [\ion{O}{III}] line measurements. 
The properties of the 62 spectra are not used in the analysis of the [\ion{O}{III}] kinematics but [\ion{O}{III}]-emission EWs are of potential interest.
As a consequence, [\ion{O}{III}] emission EWs are calculated by solving for the normalisation of a specific [\ion{O}{III}] emission `template'. 
The template is generated by running our line-fitting routine on a median composite spectrum constructed from the 268 quasars with reliable [\ion{O}{III}] line measurements. 
The spectra used to construct the composite were first de-redshifted and continuum- and \ion{Fe}{II}-subtracted.
The resulting template is mildly blue-asymmetric and has a FWHM $\simeq$900\,\kms.

The [\ion{O}{III}] model parameters and the numbers of quasars to which each model is applied are summarised in Table~\ref{tab:oiiimod}.  Example fits to spectra using the three models for [\ion{O}{III}] are shown in the top row of Fig.~\ref{fig:example_spectrum_grid}.

\begin{table}
  \centering
  \footnotesize 
    \begin{tabular}{ccc} 
    \hline
    Model & Components & Number \\
    \hline
    a & Template &  62 \\
    b & 1 Gaussian  &  128 \\
    c & 2 Gaussians &  140 \\
    \hline
    \end{tabular}
    \caption[{Summary of parameters used to fit the [\ion{O}{III}] emission, and the number of quasars to which each model is applied.}]{Summary of models used to fit the [\ion{O}{III}]\l\l4960,5008 emission, and the number of quasars to which each model is applied.}
  \label{tab:oiiimod}
\end{table} 

Figure.~\ref{fig:example_spectrum_grid} also includes example fits to eight objects spanning the full range in derived [\ion{O}{III}] properties. 
The median reduced-$\chi^2$ value for the whole sample is 1.31 and, in general, there are no significant features observable in the ${\rm spectrum}-{\rm model}$ residuals.

\subsubsection{Modelling \hans}
\label{sec:hamodel}

\begin{table}
  \centering
  \footnotesize 
    \begin{tabular}{cccc} 
    \hline
    Model     & Components & Fix centroids? & Number \\
    \hline
    a        & 1 BG  & N/A &  8 \\
    b        & 2 BG & Yes &  47 \\
    c        & 2 BG & No  &  20 \\
    d        & 2 BG + NG & Yes & $42$ \\
    e        & 2 BG + NG & No  & $48$ \\
    \hline
    \end{tabular}
    \caption[{Summary of models used to fit the \ha emission, and the number of quasars to which each model is applied.}]{Summary of models used to fit the \ha emission, and the number of quasars to which each model is applied.}
  \label{tab:hamod}
\end{table} 

There are $165$ quasars with spectra covering the \ha emission-line. 
In Appendix~\ref{sec:ch4_redshifts}, we assess the reliability of using the peak of the \ha emission as one estimate of the quasar systemic redshift 
and here we describe how the \ha emission was modelled. 

The continuum emission is first modelled and subtracted using the procedure described in \citet{coatman17}. 
We then test five different models with increasing degrees of freedom to parametrize the \ha emission. 
The models are summarised in Table~\ref{tab:hamod}. 
They are (a) a single broad Gaussian; (b) two broad Gaussians with identical velocity centroids; (c) two broad Gaussians with different velocity centroids; (d) two broad Gaussians with identical velocity centroids, and additional Gaussians to model narrow \ha emission, and the narrow components of [\ion{N}{II}]\ll$6548,6584$ and [\ion{S}{II}]\ll$6717,6731$; (e) two broad Gaussians with different velocity centroids, and additional narrower Gaussians. 
If used, the width and velocity of all narrow components are set to be equal, and the relative flux ratio of the two [\ion{N}{II}] components is fixed at the expected value of 2.96. 

In order to determine which model is selected we employ the following procedure.  
Each of the five models are fit to every spectrum and the reduced-$\chi^2$ recorded.
Initially, the model with the smallest reduced-$\chi^2$ is selected 
and the change in the reduced-$\chi^2$ measured as the complexity of the model is decreased (i.e. considering the models in Table~\ref{tab:hamod} in descending order). 
If using the simpler model results in an increase in the reduced-$\chi^2$ which is less than 10 per cent, relative to the best fitting model, then the simpler model is selected.  

\subsubsection{Emission-line parameters}

\begin{figure}
    \centering
    \includegraphics[width=\columnwidth]{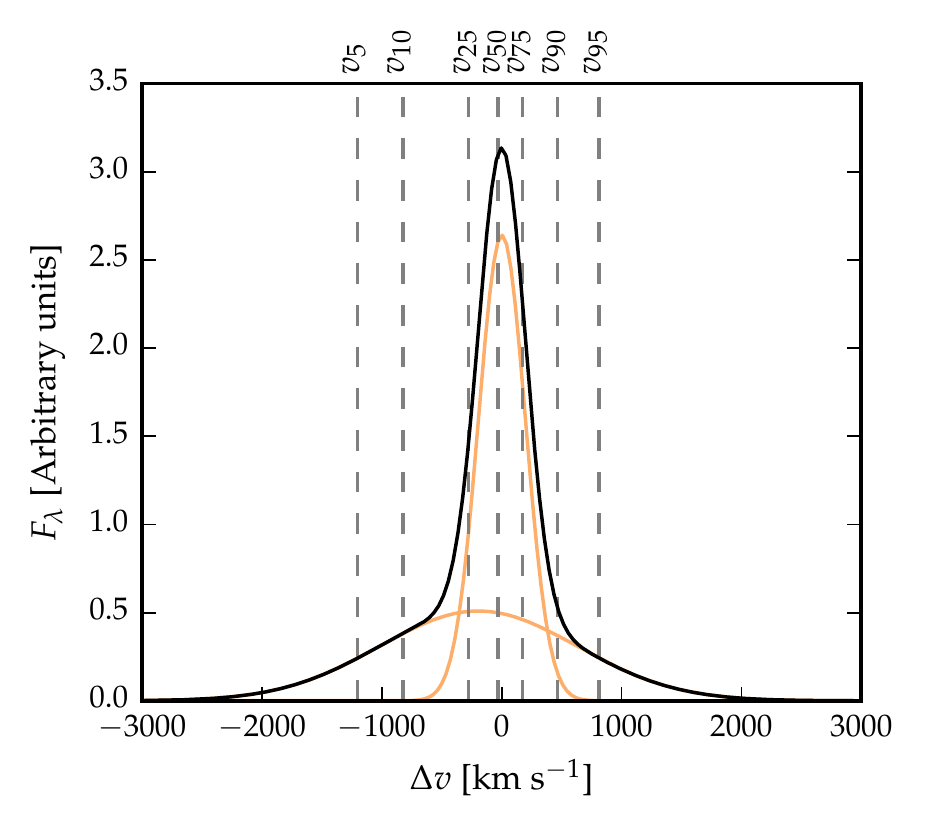} 
    \caption{The emission model, and non-parametric measures of the velocity-profile, illustrated using one component of the [\ion{O}{III}]\l\l4960,5008 emission, as described in the text.}     
    \label{fig:example_oiii}
\end{figure}

All [\ion{O}{III}] line properties are derived from the $5008.2$\,\AA\, peak but, as described above, the kinematics of [\ion{O}{III}]\l$4960.3$ are constrained to be identical. 
We do not attach any physical meaning to the individual Gaussian components used in the models. 
Decomposing the [\ion{O}{III}] emission into a narrow component at the systemic redshift and a lower-amplitude, blueshifted broad component is often highly degenerate and dependent on the spectral S/N and resolution. 
Furthermore, there is no theoretical justification that the broad component should have a Gaussian profile.  
We therefore choose to characterize the [\ion{O}{III}] line profile using a number of non-parametric measures, which are commonly used in the literature \citep[e.g.][]{whittle85,zakamska14,zakamska16}. 
The peak of the [\ion{O}{III}] emission is first defined to be at $0$\,\kms. 
A normalised cumulative velocity distribution is then constructed from the best-fitting model and the velocities below which $5$, $10$, $25$, $50$, $75$, $90$, and $95$ per cent of the total flux accumulates calculated. 
The [\ion{O}{III}] velocities for an example model are shown in Fig.~\ref{fig:example_oiii}. 

We calculate the velocity-width, $w_{90}$, containing $90$ per cent of the flux by rejecting $5$ per cent of the flux in the blue and red wings of the profile ($w_{90}\equiv v_{95} - v_{5}$).
We also calculate $w_{80}$ ($\equiv v_{90} - v_{10}$) and $w_{50}$ ($\equiv v_{75} - v_{25}$).
$w_{90}$ is relatively most sensitive to the flux in the wings of the line, whereas $w_{50}$ is relatively most sensitive to the flux in the core.  
In terms of the FWHM, $w_{50} \simeq \text{FWHM} / 1.746$, $w_{80} \simeq \text{FWHM} / 0.919$, $w_{90} \simeq \text{FWHM} / 0.716$, for a Gaussian line profile.

Employing the model-fits to the spectra we create a catalogue of derived line properties with the following fields:

\begin{itemize}
    
  \item[1] Running quasar designation. 
  
  \item[2-3] J2000.0 RA and Dec. 

  \item[4-17] $v_{5}$, $v_{10}$, $v_{25}$, $v_{50}$, $v_{75}$, $v_{90}$ and $v_{95}$ velocity of [\ion{O}{III}], relative to [\ion{O}{III}] peak, $v_{\text{peak}}$, and their errors, in \kms.  

  \item[18-19] Systemic redshift, defined using the [\ion{O}{III}] peak wavelength, and its error. 

  \item[20-25] $w_{50}$ ($\equiv v_{75} - v_{25}$), $w_{80}$ ($\equiv v_{90} - v_{10}$) and $w_{90}$ ($\equiv v_{95} - v_{5}$) velocity-width of [\ion{O}{III}], and their errors, in \kms.

  \item[26-27] Dimensionless [\ion{O}{III}] asymmetry $A$, and its error. The asymmetry is defined as 

  \begingroup\makeatletter\def\f@size{11}\check@mathfonts
   \begin{eqnarray}
    A = \frac{(v_{90} - v_{\text{peak}}) - (v_{\text{peak}} - v_{10})}{(v_{90} - v_{10})} \nonumber.     
    \end{eqnarray}  
  \endgroup

  \item[28-29] Rest-frame [\ion{O}{III}] EW, and its error, in \AA ngstroms.

  \item[30-31] [\ion{O}{III}] luminosity, and its error, in \ergs. 

  \item[32-33] $4434$-$4684$\,\AA\, rest-frame \ion{Fe}{II} EW, and its error, in \AA ngstroms.  

  \item[34-35] Velocity of \hb peak, relative to [\ion{O}{III}] peak, and its error, in \kms. 

  \item[36-37] Velocity of \ha peak, relative to [\ion{O}{III}] peak, and its error, in \kms. 

  \item[38-39] Redshift of \hb peak, and its error.

  \item[40-41] Redshift of \ha peak, and its error.

  \item[42] \ion{Fe}{II} flag. When flag is $1$ \ion{Fe}{II}-subtraction procedure has been unsuccessful (Section~\ref{sec:line_measurements}).  

  \item[43] Extreme [\ion{O}{III}] flag. When flag is $1$ [\ion{O}{III}] emission is extremely broad and blueshifted (Section~\ref{sec:extreme_oiii}). 

\end{itemize}

\subsubsection{Uncertainties on parameters}

Uncertainties on emission-line parameters derived from the best-fitting model are estimated using the Monte Carlo approach described in \citet{coatman17}. 
Two-hundred realizations of each spectrum are based on the observed quasar spectrum, with the fluxes at each wavelength generated by adding `noise', drawn from a Gaussian distribution with dispersion equal to the spectrum flux error.
The line parameter uncertainties are then estimated using half the $68$ ($84$ - $16$) percentile spread of the parameter distribution in the ensemble of simulations. 

\subsubsection{Quasars with low \rm{[O\,{\sc iii}]} EW}
\label{sec:ch4-loweqw}

\begin{figure}
    \centering
    \includegraphics[width=\columnwidth]{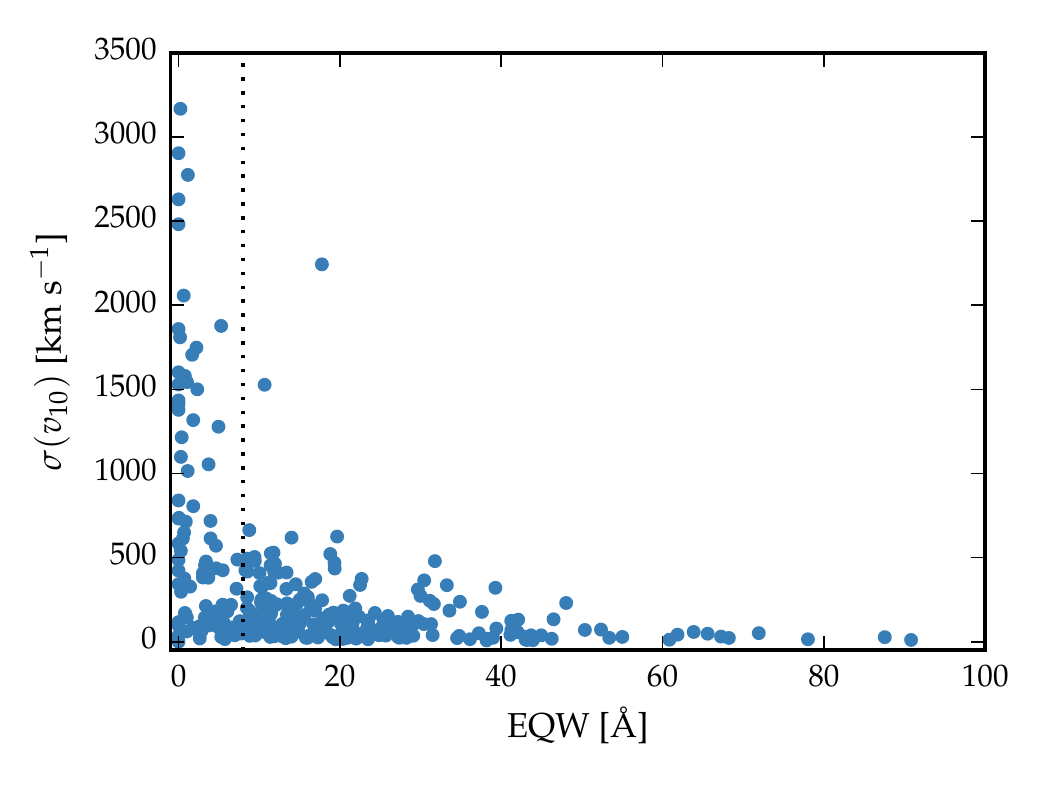} 
    \caption{Uncertainty in $v_{10}$ as a function of the EW, for [\ion{O}{III}]. Uncertainties in $v_{10}$ are large to the left of the vertical line, at $8$\,\AA. These objects are not included in our subsequent analysis of the [\ion{O}{III}] line shape.}     
    \label{fig:eqw_cut}
\end{figure}

Notwithstanding the adoption of a fixed template to describe the [\ion{O}{III}] emission in 62 of the spectra with very low [\ion{O}{III}] EWs (Section \ref{sec:oiiimodel}) the kinematic parameters for objects with low EWs are poorly constrained.
In Fig.~\ref{fig:eqw_cut} we show how the uncertainty in [\ion{O}{III}] $v_{10}$ depends on the EW. 
As the strength of [\ion{O}{III}] decreases, the average uncertainty in $v_{10}$ increases.
When the [\ion{O}{III}] $\text{EW} > 80$\,\AA, the mean uncertainty in $v_{10}$ is $50$\,\kms; increasing to $450$\,\kms\, when $10 < \text{EW} < 20$\,\AA. 
As the EW drops below $8$\,\AA, uncertainties in $v_{10}$ become very large (exceeding 1000\,\kms\, in many objects). 
[\ion{O}{III}]-parameters are thus very poorly constrained at low EW and, for the subsequent statistical investigations of the [\ion{O}{III}] kinematics, we therefore exclude objects where the measured [\ion{O}{III}] $\text{EW} < 8$\,\AA.
Application of the threshold EW leaves $226$ quasars from which kinematic information on the [\ion{O}{III}] line has been extracted. 

\section{[O\,{\sc iii}] emission properties in luminous quasars}
\label{sec:results}

\subsection{Strength and kinematics of [O\,{\sc iii}]}
\label{sec:basicresults}

\begin{figure}
    \captionsetup[subfigure]{labelformat=empty}
    \centering
    \subfloat[\label{fig:parameter_hists_a}]{}
    \subfloat[\label{fig:parameter_hists_b}]{}
    \subfloat[\label{fig:parameter_hists_c}]{}
    \subfloat[]{{\includegraphics[width=\columnwidth]{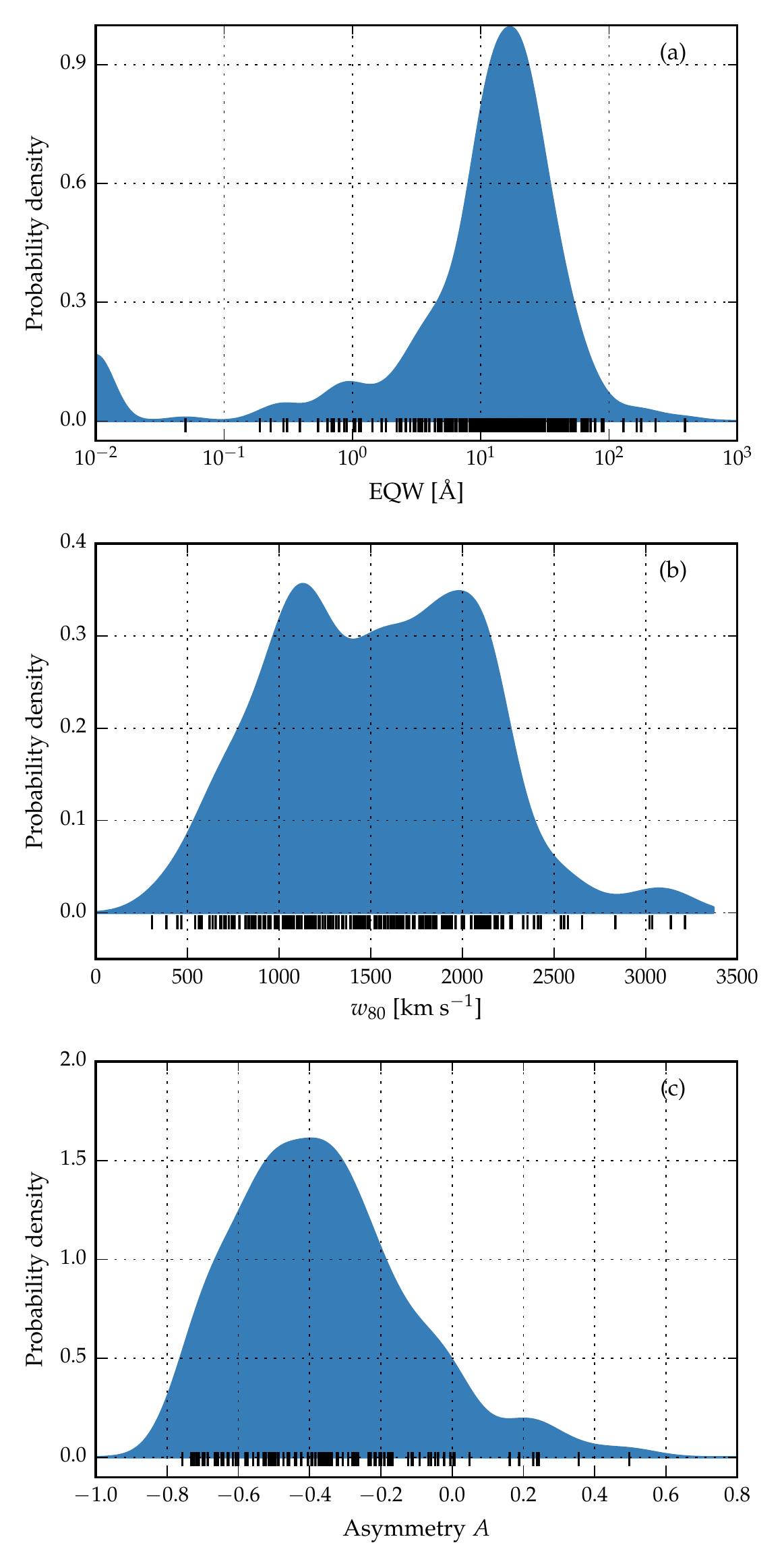} }}
\caption{Probability density distributions of the [\ion{O}{III}] parameters EW for 330 quasars (a), $w_{80}$ for 226 quasars (b) and asymmetry $A$ for 134 quasars (c), generated using Gaussian kernel density estimation. The kernel widths, which are optimised using leave-one-out cross-validation \citep{kohavi95}, are 1.4\,\AA, 140\,\kms\, and 0.08 for (a), (b) and (c) respectively. The $1200$\,\kms\, upper limit on the velocity-width of the first Gaussian function used to model [\ion{O}{III}] is responsible for the peak at $1200$\,\kms\, in (b).}
\label{fig:parameter_hists}
\end{figure}

The large sample of luminous, high-redshift broad line quasars with near infra-red spectroscopy enables, for the first time, an exploration of the full range in observed NLR properties.
The $330$ quasars exhibit a significant diversity in [\ion{O}{III}] emission properties and the probability density distribution of the [\ion{O}{III}] EW is shown in Fig.~\ref{fig:parameter_hists_a}. 
The maximum EW is $391$\,\AA, while in $10$ per cent of the sample the [\ion{O}{III}] $\text{EW} < 1$\,\AA.
The median of the distribution is $14$\,\AA\, and the $68$ percentile range is $3$ to $30$\,\AA.

For the 226 quasars where [\ion{O}{III}] EW$\ge$8\,\AA \ and the spectrum S/N allows line-widths to be measured, the median line-width (characterized by $w_{80}$ and shown in Fig.~\ref{fig:parameter_hists_b}) is $1540$\,\kms, while the $68$ percentile range is $950$ to $2100$\,\kms, with a minimum of $300$\,\kms\, and a maximum of $3200$\,\kms.

For $40$ per cent of the sample [\ion{O}{III}] is fit with a single Gaussian, where, by definition, the asymmetry is zero. 
The [\ion{O}{III}] asymmetry for the 134 quasars with two-Gaussian emission-line fits is shown in Fig.~\ref{fig:parameter_hists_c}. 
For $90$ per cent of the 134 objects [\ion{O}{III}] is blue-asymmetric.
The median asymmetry is $-0.37$ and the $68$ percentile range is $-0.61$ to $-0.12$.
Blue-asymmetric structure and high-velocity gas is generally believed to be associated with outflows. 
Our results therefore suggest that NLR outflows are prevalent in luminous high-redshift quasars. 

We also find weak correlations between these three [\ion{O}{III}] parameters. 
The EW is anti-correlated with both the line-width and asymmetry: as the [\ion{O}{III}] emission weakens it becomes broader and more blue-asymmetric, consistent with previous results \citep[e.g.][]{shen14}.  

\subsection{Luminosity-dependence of [O\,{\sc iii}] properties}
\label{sec:lumdependence}

In this section, we compare our sample of luminous $2 \lesssim z \lesssim 4$ quasars to a sample of $z\lesssim1$ SDSS quasars in order to investigate the luminosity and redshift dependence of key [\ion{O}{III}] parameters. 
We use $20\,663$ quasars with [\ion{O}{III}] measurements from the \citet{shen11} catalogue. 
The median redshift of these objects is $0.55$ and the median bolometric luminosity is $10^{45.5}$\,\ergs.

\begin{figure}
\centering 
    \includegraphics[width=\columnwidth]{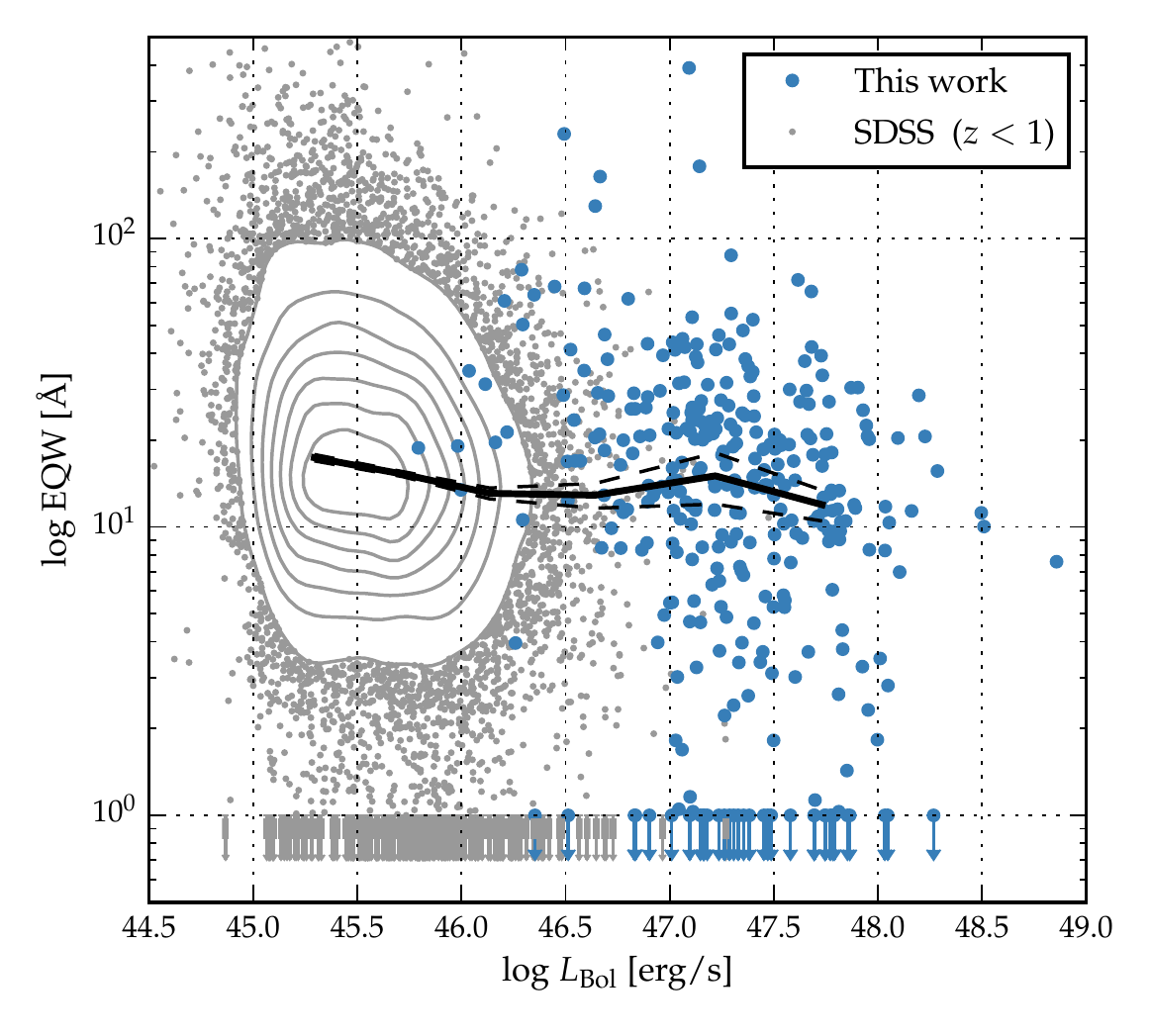} 
    \caption{The [\ion{O}{III}] EW as a function of the quasar bolometric luminosity for the luminous quasars (blue symbols) and the $z\lesssim1$ SDSS sample (grey contours and points). An upper limit at $\text{EW}=1$\,\AA\, indicates points with $\text{EW} < 1$\,\AA. The solid line shows the median [\ion{O}{III}] EW as a function of luminosity and the dashed lines show the $1$-$\sigma$ standard error on the median. The average EW decreases from $17$\,\AA\, at $L_{\text{Bol}}=10^{45.25}$\,\ergs\, to $12$\,\AA\, at $L_{\text{Bol}}=10^{47.75}$\,\ergs). The fraction of quasars with very weak [\ion{O}{III}] ($\text{EW} < 1$\,\AA) is ten times higher in the luminous quasar sample.}     
    \label{fig:eqw_lum}
\end{figure}

In Fig.~\ref{fig:eqw_lum} we show the [\ion{O}{III}] EW as a function of the quasar bolometric luminosity. 
Bolometric luminosities are estimated from monochromatic continuum luminosities at $5100$\,\AA, using the bolometric correction factor given by \citet{richards06}. 
Considering only the objects for which [\ion{O}{III}] is detected with $\text{EW} > 1$\,\AA, we observe a decrease in the [\ion{O}{III}] EW as the luminosity increases (from $17$\,\AA\, at $L_{\text{Bol}}=10^{45.25}$\,\ergs\, to $12$\,\AA\, at $L_{\text{Bol}}=10^{47.75}$\,\ergs). 
Given the luminosity spans a full $2.5$\,dex, the decrease in the [\ion{O}{III}] EW ($30$ per cent) is very modest.
However, we find [\ion{O}{III}] $\text{EW} < 1$\,\AA\, in $10$ per cent of the luminous quasars, compared to just one per cent of the $z \lesssim 1$ SDSS sample.




\begin{figure}
    \centering
    \includegraphics[width=\columnwidth]{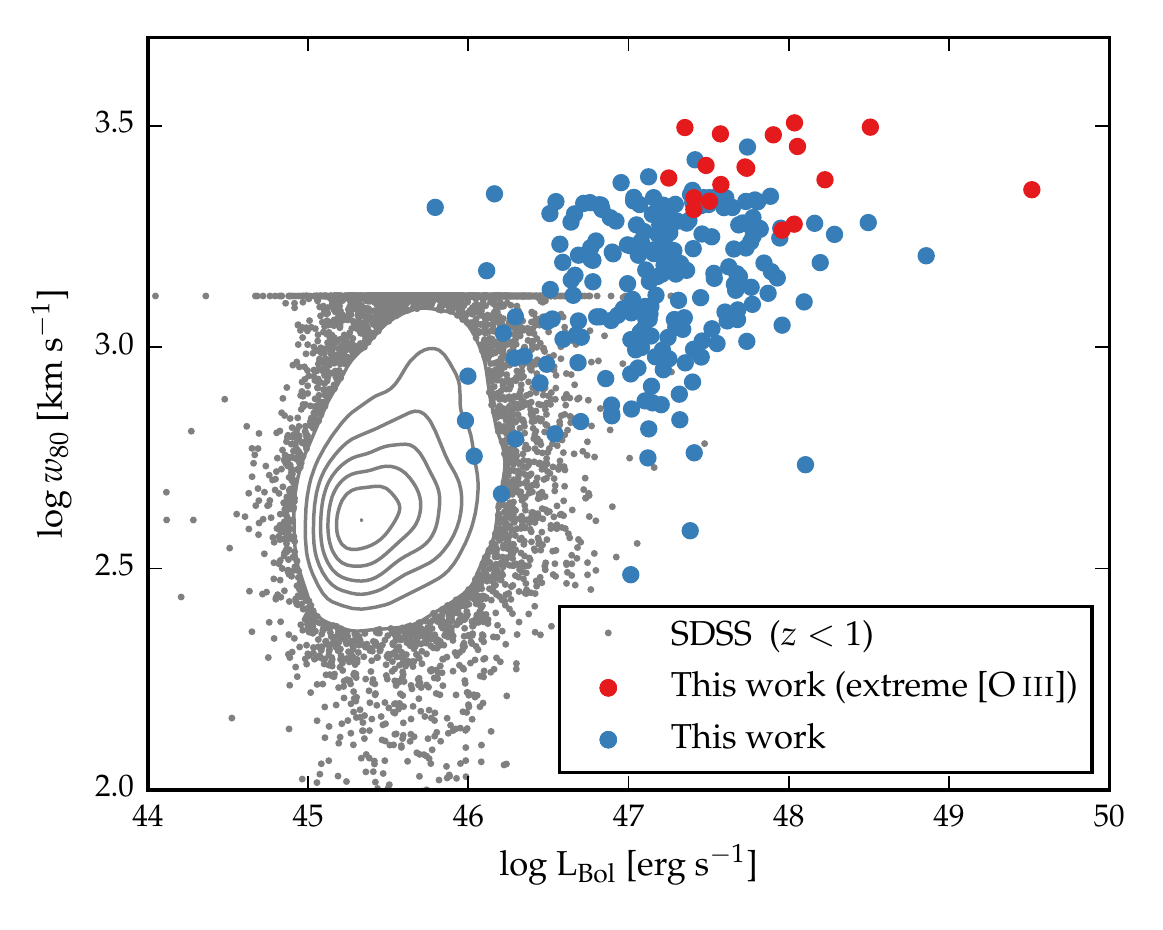} 
    \caption{[\ion{O}{III}] velocity-width, $w_{80}$, as a function of quasar bolometric luminosity. Objects with extreme [\ion{O}{III}] profiles (see Section~\ref{sec:extreme_oiii}) are shown in red. The grey contours and dots show $z\lesssim1$ SDSS DR7 quasars. The FWHM measurements given by \citet{shen11} have been converted into equivalent $w_{80}$ values by assuming $w_{80} \simeq \text{FWHM} / 0.919$. The build-up of points at $w_{80}=1300$\,\kms\, is caused by the upper-limit imposed by \citet{shen11} on the [\ion{O}{III}] FWHM. The typical [\ion{O}{III}] velocity-width increases from $440$\,\kms\, at $L_{\text{Bol}}=10^{45.5}$\,\ergs\, to $1850$\,\kms\, at $L_{\text{Bol}}=10^{48}$\,\ergs.} 
    \label{fig:lum_w80}
\end{figure}

In Fig.~\ref{fig:lum_w80} we show that the [\ion{O}{III}] velocity-width is strongly correlated with the quasar bolometric luminosity.
The typical [\ion{O}{III}] velocity-width increases from $440$\,\kms\, at $L_{\text{Bol}}=10^{45.5}$\,\ergs\, to $1850$\,\kms\, at $L_{\text{Bol}}=10^{48}$\,\ergs.  
The correlation, with the highest velocity-widths associated with the most luminous quasars, is consistent with the expectations for a model in which outflows are driven by radiative forces. 

Considering only objects in a narrow luminosity range ($10^{47} < L_{\text{Bol}} < 10^{47.5}$\,\ergs) we observe no correlations between the redshift and either the [\ion{O}{III}] velocity-width or EW.   
The lack of any evolution in typical [\ion{O}{III}] properties between $z=0$ and $z=1.5$ has previously been reported \citep[e.g.][]{harrison16}; our sample demonstrates that the [\ion{O}{III}] properties do not evolve significantly from $z=1.5$ to $z=4$.

\subsection{Relating [O\,{\sc iii}] NLR Outflow Signatures to C\,{\sc iv} BLR Outflow Signatures}
\label{sec:civtrends}

\begin{figure}
\centering 
    \includegraphics[width=\columnwidth]{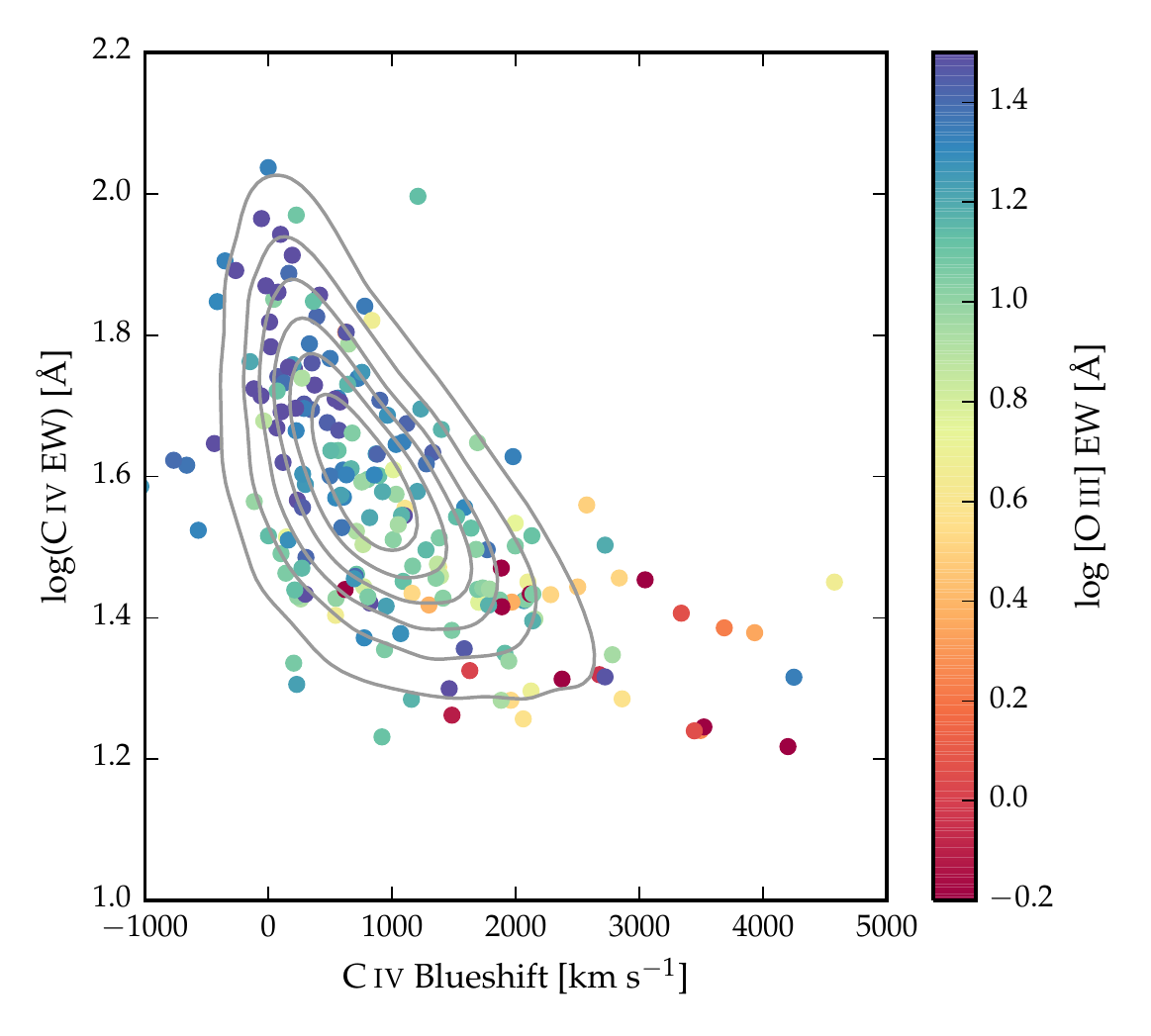} 
   \caption{Location of quasars in the \ion{C}{IV} emission-line blueshift and EW parameter space. Our sample of 213 quasars is shown as coloured circles and the locations of 32\,157 quasars from the SDSS DR7 catalogue are shown with grey contours. The [\ion{O}{III}] EW varies systematically across the parameter space. In the small \ion{C}{IV} blueshift, high EW region the mean [\ion{O}{III}] EW is $47$\,\AA, reducing to $6$\AA\, in the large \ion{C}{IV} blueshift, low EW region.}      
\label{fig:ev1} 
\end{figure}

The distribution of \ion{C}{IV} emission-line blueshift and EW provide a compact description of the diversity of broad emission-line properties in the ultraviolet for high redshift quasars \citep{sulentic07,richards11}. 
Of the 258 quasar spectra for which SDSS spectra of the \ion{C}{IV} emission are available (Section \ref{sec:data}), 45 correspond to quasars that satisfy one or more of the three criteria i) excluded from the [\ion{O}{III}]-emission sample due to poor \ion{Fe}{II} subtraction (Section \ref{sec:line_measurements}), ii) have log(\ion{C}{IV} EW)$<$1.2, and thus potentially unreliable \ion{C}{IV} emission-line blueshifts or iii) are classified as high-ionisation broad absorption line quasars. The remaining sample of 213 quasars possess both \ion{C}{IV}- and [\ion{O}{III}]-emission line EW measurements and \ion{C}{IV} blueshifts\footnote{\ion{C}{IV} emission-line parameters are measured using the prescription described in section 3.2 of \citet{coatman16}.}.

In Fig.~\ref{fig:ev1} we show the [\ion{O}{III}] EW as a function of the \ion{C}{IV} blueshift and EW.
When [\ion{O}{III}] is strong, the \ion{C}{IV} blueshift is measured using the quasar redshift calculated from the peak [\ion{O}{III}] emission. 
Otherwise, the \ion{C}{IV} blueshift is measured using the quasar redshift calculated from the \hb or \ha emission lines.  
In Appendix~\ref{sec:ch4_redshifts} we show that redshifts measured from [\ion{O}{III}], \hb and \ha are consistent to within $\sim300$\,\kms, which is small in comparison to the sample dynamic range in \ion{C}{IV} blueshift.
Also shown are contours defining the location of \ion{C}{IV} line parameters of $32\,157$ SDSS DR$7$ quasars at redshifts $1.6 < z < 3.0$. 
For the SDSS sample, systemic redshifts are taken from Allen \& Hewett (2019, in preparation). 

The [\ion{O}{III}] EW decreases systematically from the small \ion{C}{IV} blueshift, large EW region of the parameter space to the large \ion{C}{IV} blueshift, small EW region.
In the top left of the distribution (\ion{C}{IV} blueshift $<1000$\,\kms, $\text{EW} > 60$\,\AA) the mean [\ion{O}{III}] EW is $47$\,\AA, while in the bottom right, (\ion{C}{IV} blueshift $>2000$\,\kms, $\text{EW} < 30$\,\AA) the mean [\ion{O}{III}] EW is $6$\,\AA. 
For the 18 quasars with [\ion{O}{III}] $\text{EW} < 1$\,\AA \ and reliable CIV properties, the mean (median) blueshift is 2565 (2407)\,\kms, with a standard deviation of 936\,\kms.

\begin{figure}
    \centering
    \includegraphics[width=\columnwidth]{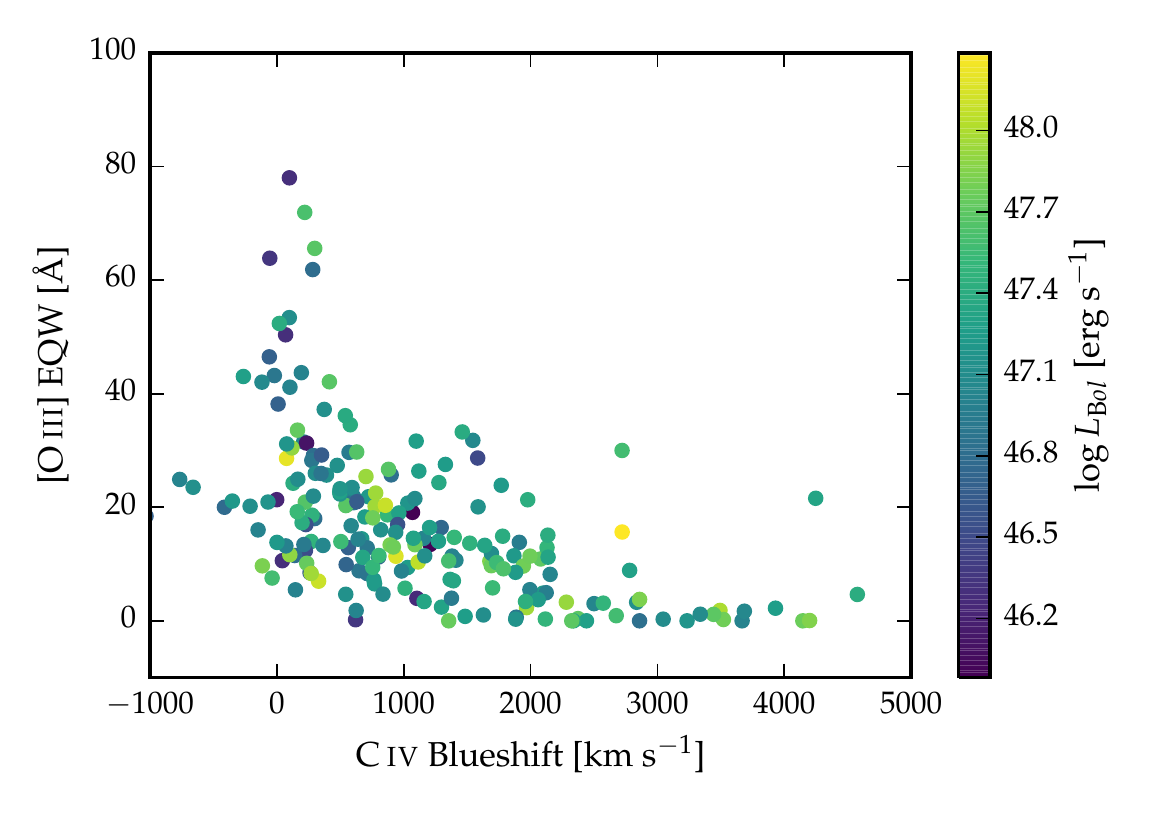} 
    \caption{[\ion{O}{III}] EW as a function of the \ion{C}{IV} blueshift for 213 quasars. The [\ion{O}{III}] EW is strongly anti-correlated with the \ion{C}{IV} blueshift. On the other hand, no strong luminosity-dependent trends (indicated by the colours of the points) are evident.}     
    \label{fig:civ_blueshift_oiii_eqw}
\end{figure}

Figure~\ref{fig:civ_blueshift_oiii_eqw} shows the [\ion{O}{III}] EW as a function of the \ion{C}{IV} blueshift for the 213 quasars.  
The luminosity of the quasars is indicated by the colour of the points. 
Both the [\ion{O}{III}] EW and the \ion{C}{IV} blueshift are known to depend on the quasar luminosity. 
However, Fig.~\ref{fig:civ_blueshift_oiii_eqw} demonstrates that the strong correlation between the \ion{C}{IV} blueshift and [\ion{O}{III}] EW is not driven by the mutual dependence of these parameters on the luminosity. 
In the context of comparison to results of other studies, it is important to note the very high luminosities of the sample, with even the faintest quasars possessing $\log L_{\text{Bol}} > 45.5$\,\ergs.

\begin{figure}
    \centering
    \includegraphics[width=\columnwidth]{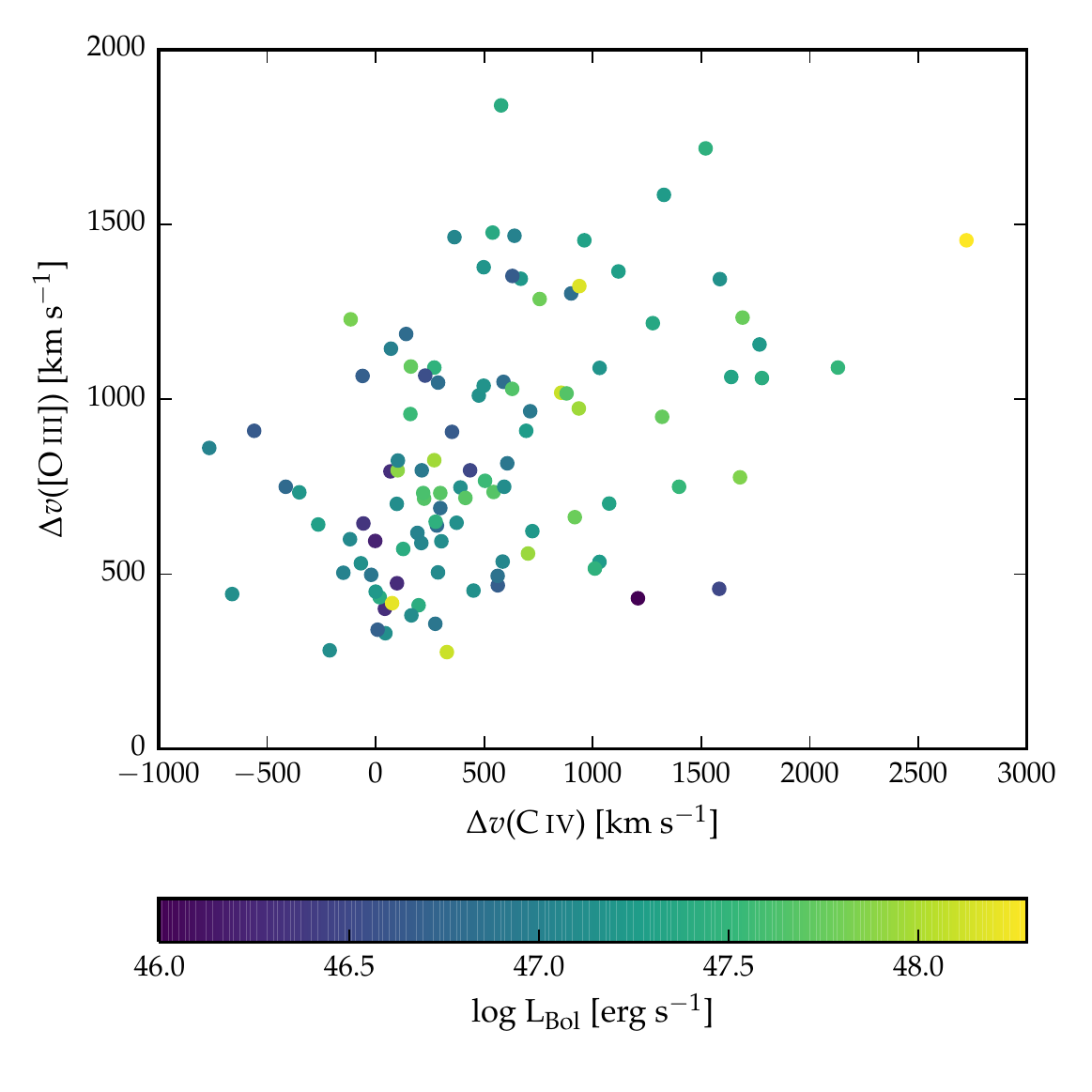} 
    \caption{Relationship between the \ion{C}{IV} ($v_{50}$(\ion{C}{IV}) - $v_{\text{peak}}$([\ion{O}{III}])) and [\ion{O}{III}] ($v_{10}$([\ion{O}{III}]) - $v_{\text{peak}}$([\ion{O}{III}])) blueshift for 107 quasars from our sample. The \ion{C}{IV} and [\ion{O}{III}] blueshifts are correlated with Spearman correlation coefficient $\rho_{\text{S}}=0.46$. The correlation is independent of the quasar luminosity (indicated by the colour of the points). }
    \label{fig:oiii_civ_blueshifts}
\end{figure}

To compare the kinematic properties of the \ion{C}{IV} and [\ion{O}{III}] emission it is necessary to further restrict the sample of quasars to ensure that the [\ion{O}{III}] emission measurements in particular are reliable. 
The [\ion{O}{III}] blueshift is defined as $v_{10}$([\ion{O}{III}]) - $v_{\text{peak}}$([\ion{O}{III}]) and the \ion{C}{IV} blueshift is defined as $v_{50}$(\ion{C}{IV}) - $v_{\text{peak}}$([\ion{O}{III}]).

Taking the sample of 213 quasars used in Figs.~\ref{fig:ev1} and \ref{fig:civ_blueshift_oiii_eqw} we therefore exclude (i) all objects with [\ion{O}{III}] EW$<$8\,\AA \ (see Section~\ref{sec:ch4-loweqw}), (ii) objects where the errors on the [\ion{O}{III}] and \ion{C}{IV} blueshifts exceed $250$ or $125$\,\kms\, respectively and (iii) two objects included among the 18 quasars with `extreme' [\ion{O}{III}]-emission (see Section~\ref{sec:extreme_oiii}). 

Figure~\ref{fig:oiii_civ_blueshifts} shows the blueshift of [\ion{O}{III}] as a function of the \ion{C}{IV} blueshift for the resulting sample of 107 quasars with high-reliability blueshift measurements. 
The dynamic range probed in this figure is limited because [\ion{O}{III}] is very weak in quasars with \ion{C}{IV} blueshifts exceeding $\sim2000$\,\kms. 
Despite this, we find the [\ion{O}{III}] blueshift to be correlated with the \ion{C}{IV} blueshift.
The Spearman correlation coefficient, $\rho_{\text{S}}$, is $0.46$, with p-value $=6\times10^{-7}$.

While deliberately having focussed on the sample of quasars with errors in blueshift measurements below $250$ or $125$\,\kms for [\ion{O}{III}] and \ion{C}{IV} respectively, 
the correlation is also evident for the sample that includes quasars with larger measurement uncertainties.  

To test the robustness of the correlation we considered a number of alternative approaches to parametrising both the [\ion{O}{III}] line shape and the systemic redshift. 
Very similar trends are observed when the [\ion{O}{III}] line shape is parametrised using $v_{25} - v_{\text{peak}}$, $v_{50} - v_{\text{peak}}$, $w_{80} = v_{90} - v_{10}$, or the asymmetry $A$.
The same trend is also observed when the systemic redshift is defined using the peak of the \hb emission. 

The existence of a correlation between the [\ion{O}{III}] and \ion{C}{IV} blueshift shows that high-velocity winds in the NLR are preferentially seen when strong winds are driven in the vicinity of the central engine.
Figure~\ref{fig:oiii_civ_blueshifts} demonstrates that this correlation is not driven by the luminosity. 



\subsection{Extreme [O\,{\sc iii}] profiles}
\label{sec:extreme_oiii}

\begin{figure*}
    \centering
    \includegraphics[width=\textwidth]{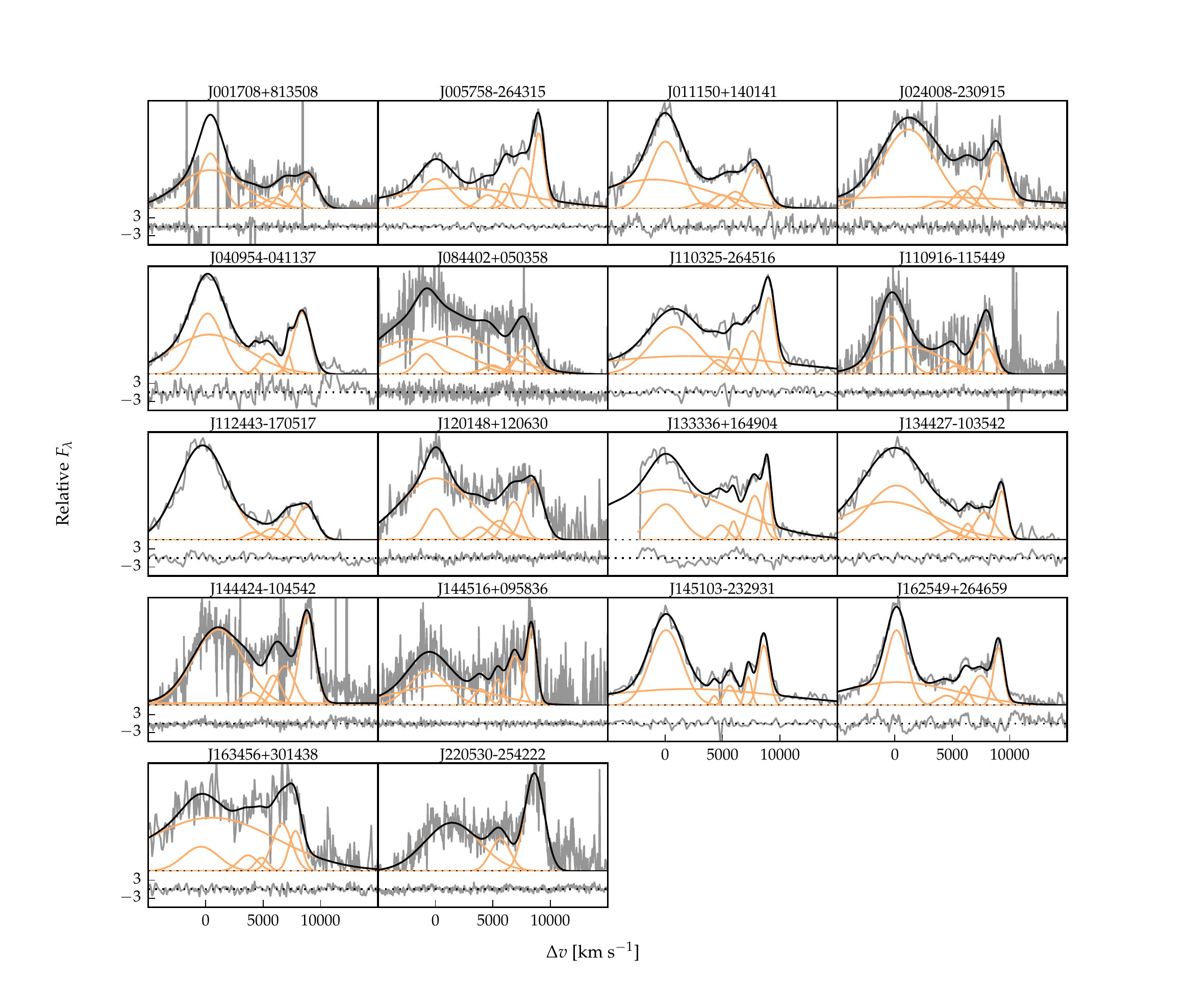} 
    \caption{Model fits to the continuum- and \ion{Fe}{II}-subtracted \hbns/[\ion{O}{III}] emission in $18$ quasars with extreme [\ion{O}{III}] emission profiles. The data is shown in grey, the best-fitting model in black, and the individual model components in orange. The peak of the [\ion{O}{III}] emission is used to define the systemic redshift, and $\Delta{v}$ is the velocity shift from the rest-frame transition wavelength of \hbns. Below each spectrum, the data- minus-model residuals, scaled by the errors on the fluxes, are shown.}     
    \label{fig:example_spectrum_grid_extreme_oiii}
\end{figure*}


Figure~\ref{fig:example_spectrum_grid_extreme_oiii} shows the spectra of $18$ objects which we visually identified as having [\ion{O}{III}] emission profiles with similar characteristics to four extremely dust-reddened quasars at $z\sim2$ identified by \citet{zakamska16}. 
The extreme nature of the [\ion{O}{III}] emission in their sample of red quasars led \citet{zakamska16} to propose that these objects are observed in the process of expelling the gas in their host-galaxies and transitioning from a dust-obscured, starburst phase to a luminous, blue quasar \citep[e.g.][]{sanders88}.

The [\ion{O}{III}] emission in the $18$ objects in our sample is very broad ($1800 \lesssim w_{80} \lesssim 3200$\,\kms; Fig.~\ref{fig:lum_w80}). 
In many of these objects the systemic, core component of [\ion{O}{III}] is not detected.
The [\ion{O}{III}] doublet is blended together, and is also heavily blended with the red wing of the \hb emission. 
Flux from additional ions, such as \ion{Fe}{II} emission, may also be making a significant contribution in the 4800-5000\,\AA \ region of the spectra. 

\begin{figure}
\centering 
    \includegraphics[width=\columnwidth]{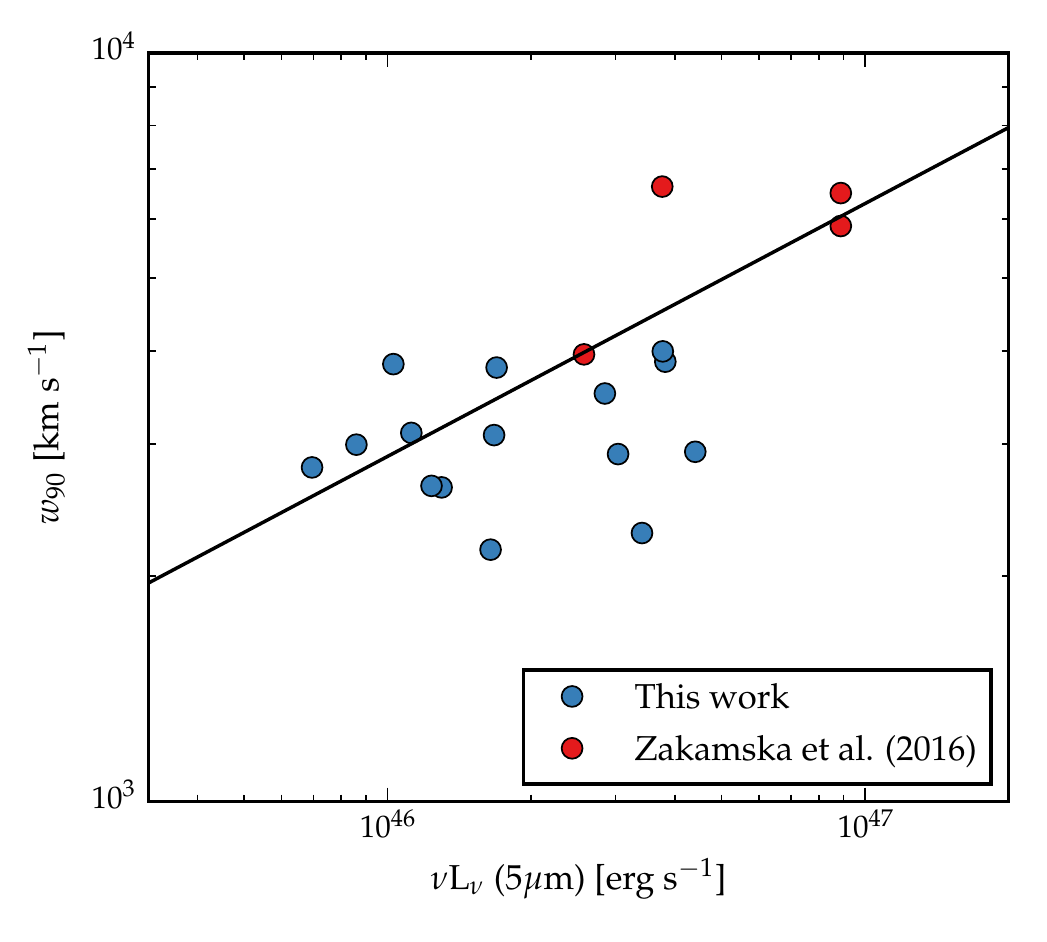} 
    \caption{Velocity-widths, $w_{90}$, and rest-frame $5$\,$\mu$m luminosities of the $18$ quasars in our sample with extreme [\ion{O}{III}] emission profiles and the four dust-reddened quasars from \citet{zakamska16}. Our sample lies close to the width-luminosity relation derived by Zakamska et al.}     
    \label{fig:fivemicron_w90}
\end{figure}

In Fig.~\ref{fig:fivemicron_w90} we compare the velocity-widths and rest-frame 5\,$\mu$m luminosities of the 18 quasars in our sample with the four quasars from \citet{zakamska16}.
On average, the \citeauthor{zakamska16} quasars have higher luminosities ($10^{46.7}$ versus $10^{46.3}$\ergs) but the difference is not unexpected given that the Zakamska-sources were selected to have very red $(r-W4)$ colours. 
The [\ion{O}{III}] emission in the \citeauthor{zakamska16} quasars is also broader than in our sample ($w_{90}=5740$ versus 3120\,\kms).
However, the [\ion{O}{III}] velocity-width and luminosity of the least extreme \citeauthor{zakamska16} quasar has properties very similar to our sample, and our objects are consistent with the $w_{90}$-5\,$\mu$m luminosity relation derived by \citeauthor{zakamska16} (Fig.~\ref{fig:fivemicron_w90}).
While certainly relatively rare, quasars with extreme [\ion{O}{III}] line properties similar to the red quasars studied by \citeauthor{zakamska16} are present in the population of luminous, high-redshift quasars with conventional ultraviolet SEDs.


\section{Discussion}
\label{sec:discussion}

The object sample used here is unique, consisting of a large number of high-luminosity quasars drawn from the population of `normal' objects. 
The investigation thus compliments the extensive work that has focussed on reddened, or obscured, AGN \citep[e.g.][and references therein]{dipompeo18,sun18}.

The three main observational results presented in Section \ref{sec:results} are i) the correlation between the blueshifts of the [\ion{O}{III}] emission and the \ion{C}{IV} emission, ii) the anti-correlation between the [\ion{O}{III}]-emission EW and the blueshift of the \ion{C}{IV} emission and iii) the order of magnitude increase in the fraction of quasars which exhibit undetectable [\ion{O}{III}] emission for the high-redshift, high-luminosity ($\gtrsim 10^{47.0}$\ergs) objects compared to the SDSS-quasars at lower redshifts and luminosities ($\sim10^{45.5}$\ergs). 

Blueshifted \ion{C}{IV} emission is thought to arise in a high-velocity accretion disc wind \citep[see][for a recent review]{netzer15} with much of the emitting material believed to exist at small distances, $r \lesssim 1$\,pc, from the supermassive black hole. 
Detections of [\ion{O}{III}] emission on scales of many kiloparsecs have been reported from studies of relatively small sample of lower luminosity AGN \citep[e.g.][]{liu13, harrison14} but the location of the [\ion{O}{III}]-emitting gas in the high-luminosity sample investigated here is not known. Indeed, a potential physical cause of the change in the [\ion{O}{III}] emission strength, particularly the reduction in the core-component, is a systematic change in the quasar ultra-violet/X-ray SED, resulting in photo-ionisation-induced variations in the strength of the NLR emission.

The implications of the results in Section \ref{sec:results} for our understanding of the effect of outflowing material from luminous quasars on their host galaxies therefore depends on the physical location of the [\ion{O}{III}]-emitting gas and a better understanding of the impact of photo-ionisation effects.

A significant step forward would be the unambiguous detection of extended [\ion{O}{III}] emission and a reliable distance determination for the emitting gas. While integral field unit observations provide spatially-resolved observations, the high-luminosity of the quasars and redshifts of $z\simeq2 - 3$ make such observations extremely challenging. The difficulties are illustrated by the caution expressed about the inferred spatial extent of [\ion{O}{III}] emission in a number of studies of luminous quasars due to the need to model the point spread function with high accuracy \citep{husemann16}.

Notwithstanding the lack of unambiguous evidence linking the small-scale BLR outflows with [\ion{O}{III}]-emitting gas at distances of kilo-parsecs in our high-luminosity quasars, such a possibility has received recent attention \citep[e.g.][]{zakamska16}. In such a scenario the strong anti-correlation between the \ion{C}{IV} blueshift and the [\ion{O}{III}] EW suggests that outflows are having a significant impact on gas extended over kilo-parsec scales in the NLR.
Dynamical time-scales for the impact of fast moving outflows even on large NLRs are very short: it would take only $10^6$ years for an outflow travelling at $3000$\,\kms\, to reach $3$\,kpc. 
Lifetimes of luminous quasars at these redshifts may be $\sim 10^7$ years \citep[e.g.][]{martini01} and if the BLR outflows can break out into the interstellar medium of the host-galaxy, the NLR could be cleared on a relatively short time-scale.
One possibility is that the BLR winds collide with the interstellar medium, shocking and accelerating material to produce a galaxy-wide wind \citep[e.g.][]{king11,faucher12}. 
As \ion{C}{IV} blueshifts are generally weaker in lower-luminosity quasars 
such a model also explains our finding that objects with very weak [\ion{O}{III}] ($\text{EW} < 1$\,\AA) are ten times rarer in $z \lesssim 1$ SDSS quasars than in our sample of very luminous quasars.   

\subsection{Possible mass outflow rate and kinetic power}

The mass outflow rate and kinetic power of galaxy-wide outflows are important properties in order to understand the role played by quasars in the evolution of galaxies. 
Under the assumption that the results presented in Section \ref{sec:results} do arise as the result of outflows affecting a NLR extending to kiloparsec scales we calculate the mass outflow rate ($\dot{M}$) and kinetic power ($P_{\text{K}}$) of the ionised outflows using the [\ion{O}{III}] emission as a gas tracer \citep[e.g.][]{harrison12,cano-diaz12,liu13,brusa15,carniani15,bischetti16,kakkad16}.  
By making reasonable assumptions for unknown quantities (including the geometry, spatial scale and density of the gas in the outflow) we can calculate order of magnitude estimates of the outflow properties.
Our calculations are based on the model of \citet{cano-diaz12}, and a comprehensive description of the assumptions in the model and their impact on the inferred outflow properties can be found in \citet{cano-diaz12} and \citet{kakkad16}. 

Following \citeauthor{cano-diaz12}, the mass in the ionised outflow is given by 

\begin{eqnarray}
M \simeq 5.33 \times 10^7 \left( \frac{C}{10^{[\text{O/H}] - [\text{O/H}]_\odot}} \right) \left( \frac{L([\text{O}\,\textsc{iii}])}{10^{44}\, \text{erg}\,\text{s}^{-1}}\right) \nonumber \\ \times \left\langle \frac{n_e}{10^3\, \text{cm}^{-3}} \right\rangle^{-1} \, M_\odot
\end{eqnarray}

\noindent where $L([\text{O}\,\textsc{iii}])$ is the luminosity of [\ion{O}{III}] emitted in the outflow (in units of $10^{44}$\,\ergs), $\langle n_e \rangle$ is the electron density in the outflowing gas (in units of $10^3$\,cm$^{-3}$), $10^{[\text{O/H}]}$ is the metallicity (in units of Solar metallicity), $C$ ($=\langle n_e \rangle ^2 / \langle n_e^2\rangle$) is the condensation factor (taken to be $\simeq1$). 
Assuming a conical outflow with uniformly distributed clouds out to a radius $R$ with a constant outflow velocity, the mass outflow rate of the gas is given by

\begin{eqnarray}
\dot{M} = 164 \left( \frac{R}{1\,\text{kpc}} \right)^{-1} \left( \frac{C}{10^{[\text{O/H}] - [\text{O/H}]_\odot}} \right) \left( \frac{L([\text{O}\,\textsc{iii}])}{10^{44}\, \text{erg}\,\text{s}^{-1}}\right) \nonumber \\ \times \left( \frac{v}{1000\,\text{km}\,\text{s}^{-1}}\right) \times \left\langle  \frac{n_e}{10^3\, \text{cm}^{-3}} \right\rangle^{-1} \, M_\odot \, \text{yr}^{-1}
\end{eqnarray}

\noindent where $v$ is the outflow velocity (in units of $1000$\,\kms).
The kinetic power of the outflow ($1/2\dot{M}v^2$) is given by: 

\begin{eqnarray}
P_{\text{K}} = 5.17 \times 10^{43} \left( \frac{R}{1\,\text{kpc}} \right)^{-1} \left( \frac{C}{10^{[\text{O/H}] - [\text{O/H}]_\odot}} \right) \left( \frac{L([\text{O}\,\textsc{iii}])}{10^{44}\, \text{erg}\,\text{s}^{-1}}\right) \nonumber \\ \times \left( \frac{v}{1000\,\text{km}\,\text{s}^{-1}}\right)^3 \times \left\langle \frac{n_e}{10^3\, \text{cm}^{-3}} \right\rangle^{-1} \, \text{erg}\,\text{s}^{-1}
\end{eqnarray}

\noindent We assume the outflowing gas is represented by the broader of the two Gaussian components in our [\ion{O}{III}] model and use the luminosity of this component to estimate the luminosity of the outflowing gas. 
The maximum outflow velocity ($\simeq v_{5}$ -- see Fig.\ref{fig:example_oiii}) is adopted as representative of the average outflow velocity, with the lower velocities due to projection effects \citep{cano-diaz12}.
When the [\ion{O}{III}]-emission is reproduced by a single Gaussian there is no detectable outflow component and such objects are not considered here. 
For the outflow radius we use 4\,kpc, which is broadly consistent with spatially resolved observations of quasars at similar redshifts and luminosities \citep[e.g.][]{cano-diaz12,carniani15,brusa16} and photo-ionisation estimates \citep[e.g.][]{zakamska16} but see \citet{husemann16} for a different view.

With these values, we calculate outflow rates which range from a few to $4000\,M_\odot\,\text{yr}^{-1}$. 
The mean for our sample of 134 objects is $560\,M_\odot\,\text{yr}^{-1}$. 
The kinetic power of the outflows ranges from $\simeq10^{41.8}$ to $\simeq10^{45.7}$\,\ergs, with a mean of $10^{44.7}$\,\ergs. 
The mean kinetic power corresponds to $\sim0.15$ per cent of the bolometric luminosity, reaching almost 1 per cent in the most powerful outflows.  
However, if the ionised outflow is accompanied by a neutral/molecular outflow an order of magnitude more massive, then the kinetic power is also likely to be an order of magnitude higher \citep{cano-diaz12}, i.e. about $1.5$ per cent of the bolometric luminosity for the mean kinetic power and 10 per cent for the most powerful outflows. 
Such outflow efficiencies are in accord with recent AGN feedback models \citep[e.g.][]{zubovas12}, which predict a coupling efficiency between AGN-driven outflows and AGN power of $\sim5$ per cent (see \citet{king15} for a review of the fundamental theoretical considerations).

\subsection{Comparison to established [\ion{O}{\sc III}] emission trends within the quasar population}

The systematic, albeit with a very large scatter, decrease in the [\ion{O}{III}] EW with quasar luminosity shown in Fig.~\ref{fig:eqw_lum} is in agreement with a number of previous investigations \citep[e.g.][]{brotherton96,sulentic04,baskin05b,zhang11,stern12}.
The origin of this correlation -- known as the [\ion{O}{III}] Baldwin effect \citep[e.g.][]{baldwin77} -- has not been demonstrated conclusively. 
The size of the NLR (and hence the [\ion{O}{III}] luminosity) is predicted to scale with the square root of the luminosity of the source of ionising photons \citep[e.g.][]{netzer90} and low-luminosity Seyfert galaxies appear to obey such a relationship \citep[e.g.][]{bennert02}. 
Extrapolating to high luminosity quasars leads to the prediction of NLRs with galactic dimensions.
Under these conditions, the size of the NLR will be limited by the density and ionisation state in the NLR. 
In other words, the NLR cannot continue to grow beyond the radius at which there is no longer gas available to be ionised and the luminosity of the NLR is predicted to saturate \citep[e.g.][]{hainline13,hainline14}. 

\citet{shen14} showed how the [\ion{O}{III}] EW decreases as the optical \ion{Fe}{II} strength (which is related to the Eddington ratio) or luminosity increase. 
However, the amplitude of the systemic, core [\ion{O}{III}] emission decreases faster than the wing component.
A by-product of the reduction of the line core strength is that, overall, the [\ion{O}{III}] profile becomes broader and more blueshifted as the broad wing becomes relatively more prominent.
If the anti-correlation between the [\ion{O}{III}] EW and \ion{C}{IV} blueshift is primarily driven by a reduction in the flux of the core component (as a stable NLR is removed by the outflowing material), a correlation between the [\ion{O}{III}] blueshift and \ion{C}{IV} blueshift, similar to the one seen in Fig.~\ref{fig:oiii_civ_blueshifts}, is expected. 
A reduction in the flux of the core component could also explain the anti-correlation between the [\ion{O}{III}] EW and blue-asymmetry / velocity-width reported in Section~\ref{sec:basicresults}. 

\section{Conclusions}

We have assembled rest-frame ultraviolet and and optical spectra of 330 high luminosity, $45.5 \lesssim \log L_{\text{Bol}} \lesssim 49.0$\,\ergs, quasars with redshifts $1.5 \lesssim z \lesssim 4.0$. For the first time, the sample allows an investigation of the \ion{C}{IV}\ll 1548,1550 emission and [\ion{O}{III}]\ll 4960,5008 emission properties of high-luminosity, but otherwise `normal', quasars.
The main observational results are:

\begin{itemize}

\item The correlation between the blueshifts of the [\ion{O}{III}] emission and the \ion{C}{IV} emission.

\item The anti-correlation between the [\ion{O}{III}]-emission EW and the blueshift of the \ion{C}{IV} emission.

\item The order of magnitude increase in the fraction of quasars which exhibit undetectable [\ion{O}{III}] emission for the high-redshift, high-luminosity ($\gtrsim 10^{47.0}$\ergs) objects compared to SDSS-quasars at lower redshifts and luminosities ($\sim10^{45.5}$\ergs). 

\item Eighteen objects possess very broad [\ion{O}{III}] emission profiles with similar characteristics to four extremely dust-reddened quasars recently identified by \citet{zakamska16}. The properties of the 18 objects demonstrate that such high-velocity [\ion{O}{III}] emission is present in a fraction of otherwise normal high-luminosity quasars.

\end{itemize}

The implications of the results presented in Section \ref{sec:results} for our understanding of the effect of outflowing material from luminous quasars on their host galaxies depends on the physical location of the [\ion{O}{III}]-emitting gas. If the [\ion{O}{III}]-emitting gas is located on physical scales of a kilo-parcsec or more from the quasar, then the observations provide direct evidence for the impact of quasar activity on galactic scales. Such a conclusion, however, is dependent on yet to be obtained direct evidence that the [\ion{O}{III}]-emitting gas in such high-luminosity quasars extends to kilo-parsec scales.
 
\section*{Acknowledgements}

We thank the referee, Yue Shen, for a constructive review that resulted in an improved presentation of the results. LC thanks the Science and Technology Facilities Council (STFC) for the award of a studentship. MB acknowledges support from STFC via an Ernest Rutherford Fellowship and the Royal Society via a University Research Fellowship. PCH acknowledges support from the STFC via a Consolidated Grant to the Institute of Astronomy, Cambridge.




\bibliographystyle{mnras}
\bibliography{bibliography}



\appendix

\section{Reliability of systemic redshift estimates}
\label{sec:ch4_redshifts}

\begin{figure}
   \captionsetup[subfigure]{labelformat=empty}
    \centering
    \subfloat[\label{fig:redshift_comparison_a}]{}
    \subfloat[\label{fig:redshift_comparison_b}]{}
    \subfloat[\label{fig:redshift_comparison_c}]{}
    \subfloat[]{{\includegraphics[width=\columnwidth]{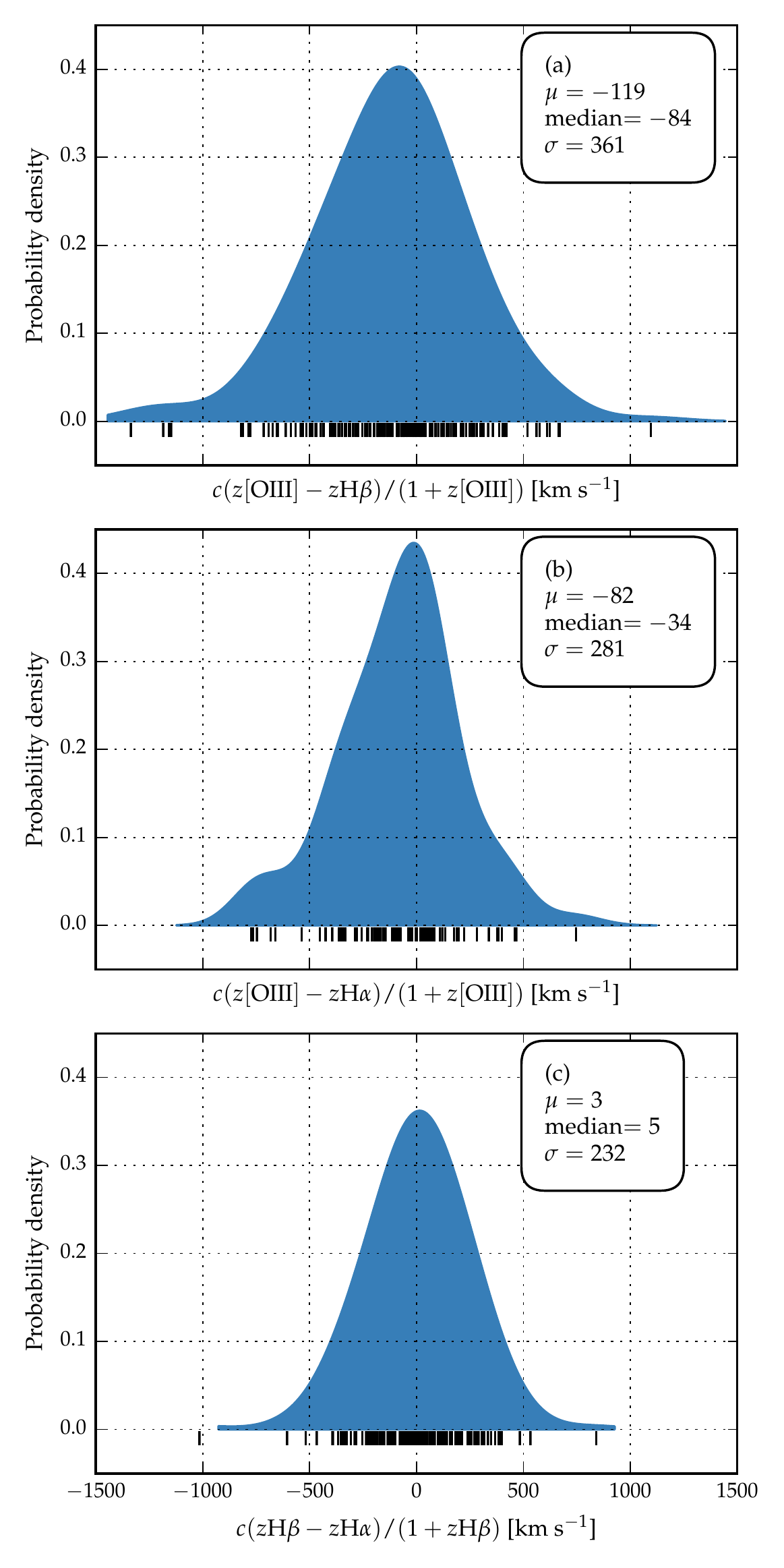} }}
    \caption{Comparison of systemic redshift estimates using [\ion{O}{III}], \hb and \hans. The probability density distributions are generated using a Gaussian kernel density estimator with $170$, $120$ and $140$\,\kms\, kernel widths for (a), (b) and (c) respectively. The short black lines show the locations of the individual points.}       
    \label{fig:redshift_comparison}
\end{figure}

In this section, we compare systemic redshift estimates based on [\ion{O}{III}], \hb and \hans. 
Redshifts are calculated using the wavelength of the peak flux in the best-fitting parametric model, which, in the case of the Balmer lines, may include broad and narrow emission components. 

We compare systemic redshift estimates based on [\ion{O}{III}] and \hb (Fig.~\ref{fig:redshift_comparison_a}), [\ion{O}{III}] and \ha (Fig.~\ref{fig:redshift_comparison_b}) and \hb and \ha (Fig.~\ref{fig:redshift_comparison_c}). 
We generate probability density distributions using a Gaussian kernel density estimator.
The kernel width, which is optimised using leave-one-out cross-validation, is $170$, $120$ and $140$\,\kms\, in Fig.s~\ref{fig:redshift_comparison_a}, \ref{fig:redshift_comparison_b} and \ref{fig:redshift_comparison_c} respectively. 

There are $182$, $85$ and $162$ objects compared in Fig.s~\ref{fig:redshift_comparison_a}, \ref{fig:redshift_comparison_b} and \ref{fig:redshift_comparison_c} respectively. 
We have excluded [\ion{O}{III}], \hb and \ha measurements when the uncertainties on the peak velocities exceed $200$, $300$ and $200$\,\kms\, respectively. 
We also exclude [\ion{O}{III}] measurements from 16 objects with very broad, blueshifted [\ion{O}{III}] emission that is strongly blended with the red wing of \hb (these objects are discussed in Section~\ref{sec:extreme_oiii}).

The scatter between the different redshift estimates ($360$, $280$, and $230$\,\kms\, in Figures~\ref{fig:redshift_comparison_a}, \ref{fig:redshift_comparison_b} and \ref{fig:redshift_comparison_c} respectively) is consistent with previous studies of redshift uncertainties from broad emission-lines \citep[e.g.][]{shen16b}. 
The systematic offset between the \ha and \hb estimates is effectively zero. 
However, the [\ion{O}{III}] redshifts appear to be systematically offset in comparison to both \ha and \hbns, in the sense that [\ion{O}{III}] is blueshifted in the rest-frame of the Balmer lines. 
This effect is strongest when [\ion{O}{III}] is compared to \hbns, in which case [\ion{O}{III}] is shifted by $\sim100$\,\kms\, to the blue.

\citet{hewett10} found that [\ion{O}{III}] was blueshifted by $\sim45$\,\kms\, relative to a rest-frame defined using photospheric \ion{Ca}{II}\ll$3935$,$3970$ absorption in the host galaxies of $z<0.4$ SDSS AGN and that [\ion{O}{III}] is increasingly blue-asymmetric at higher luminosities. 
Therefore, the $100$\,\kms offset we measure is consistent with \citet{hewett10} once the very different luminosities of the two samples are accounted for. 


\bsp	
\label{lastpage}
\end{document}